\documentclass[a4paper,11pt, english]{article} 
\usepackage{ pool/jinstpub}
\usepackage{babel}
\usepackage{graphbox}
\usepackage{multirow}
\usepackage{makecell}
\usepackage{amsmath,amssymb, gensymb}
\usepackage{natbib,hyperref}
\usepackage{url}
\usepackage{siunitx}
\usepackage{xspace}
\usepackage{microtype}
\usepackage{subfig}
\usepackage[useregional]{datetime2}
\usepackage{graphicx}

\usepackage{siunitx}
\usepackage{xspace}

\newcommand{\NEW}{NEXT-White}
\newcommand{\NEXT}{NEXT-100}

\newcommand{\fig}{figure}

\newcommand{\Fig}{Figure}

\newcommand{\Tab}{Table}

\newcommand{\ie}{{\it i.e.}}

\newcommand{\micro}{\ensuremath{\mu}}
\newcommand{\chitwo}{\ensuremath{\chi^2}}

\newcommand{\stat}{\textrm{(stat.)}}
\newcommand{\sys}{\textrm{(sys.)}}
\newcommand{\bbtnu}{\ensuremath{\beta\beta2\nu}}
\newcommand{\tz}{\ensuremath{t_0}}
\newcommand{\st}{\ensuremath{S_2}}
\newcommand{\so}{\ensuremath{S_1}}

\newcommand{\Qbb}{\ensuremath{Q_{\beta\beta}}}

\newcommand{\Kr}[1]{\ensuremath{^{#1}\mathrm{Kr}}\xspace}

\newcommand{\Rb}[1]{\ensuremath{^{#1}\mathrm{Rb}}\xspace}

\DeclareSIUnit\c{\mbox{$c$}}
\DeclareSIUnit\magn{\mbox{$\times$}}
\DeclareSIUnit\min{min}
\DeclareSIUnit\week{week}
\DeclareSIUnit\year{yr}
\DeclareSIUnit\years{years}
\DeclareSIUnit\yr{yr}
\DeclareSIUnit\standard{std}
\DeclareSIUnit\str{sr}
\DeclareSIUnit\ppm{ppm}
\DeclareSIUnit\ppb{ppb}
\DeclareSIUnit\ppt{ppt}
\DeclareSIUnit\pe{PE}
\DeclareSIUnit\spe{SPE}
\DeclareSIUnit\ev{events}
\DeclareSIUnit\ct{counts}
\DeclareSIUnit\neutron{\mbox{$n$}}
\DeclareSIUnit\smp{samples}
\DeclareSIUnit\Sample{S}
\DeclareSIUnit\ch{ch}
\DeclareSIUnit\hit{hit}
\DeclareSIUnit\hits{hits}
\DeclareSIUnit\bin{(\mbox{5-PE}~bin)}
\DeclareSIUnit\sgm{\mbox{$\sigma$}}
\DeclareSIUnit\rms{RMS}
\DeclareSIUnit\keVr{\mbox{keV$_{\rm nr}$}}
\DeclareSIUnit\keVee{\mbox{keV$_{e{\rm e}}$}}
\DeclareSIUnit\ph{photon}
\DeclareSIUnit\pes{pes}
\DeclareSIUnit\el{electrons}
\DeclareSIUnit\pm{PMT}
\DeclareSIUnit\inch{"}
\DeclareSIUnit\bit{bit}
\DeclareSIUnit\sample{samples}
\DeclareSIUnit\barn{barn}
\DeclareSIUnit\bara{bar}
\DeclareSIUnit\barg{barg}
\DeclareSIUnit\mlardepth{\mbox(meter~of~\LAr~depth)}
\DeclareSIUnit\Curie{Ci}
\DeclareSIUnit\psi{psi}
\DeclareSIUnit\parsec{pc}
\DeclareSIUnit\liveday{\mbox{live-days}}
\DeclareSIUnit\days{\mbox{days}}
\DeclareSIUnit\day{\mbox{day}}
\DeclareSIUnit\miles{\mbox{miles}}
\DeclareSIUnit\degreeC{\mbox{$^{\circ}$C}}
\DeclareSIUnit\electron{\mbox{$e^-$}}
\DeclareSIUnit\Euro{\mbox{\euro}}
\DeclareSIUnit\cph{cph}
\DeclareSIUnit\neq{neq}
\DeclareSIUnit\unit{unit}
\DeclareSIUnit\byte{Byte}
\DeclareSIUnit\Bq{\becquerel}

\newcommand{\XeWaveLength}{\SI{172}{\nano\meter}}

\newcommand{\HPXeEL}{HPXe-EL}

\newcommand{\RII}{Run II}

\newcommand{\XenonFanoFactor}{\num{0.15}}
\newcommand{\XenonEnergyPerElectron}{\SI{21.9}{\eV}}
\newcommand{\XenonQbb}{\SI{2458}{\keV}}
\newcommand{\XenonElectronsAtQbb}{\SI{112237}{\el}}

\newcommand{\LT}{\ensuremath{\tau}}
\newcommand{\SQRE}{\ensuremath{1/\sqrt{E}}}

\newcommand{\X}{\ensuremath{x}}
\newcommand{\R}{\ensuremath{r}}
\newcommand{\Y}{\ensuremath{y}}
\newcommand{\Z}{\ensuremath{z}}

\newcommand{\VD}{\ensuremath{v_d}}

\newcommand{\XY}{\ensuremath{(x, y)}}
\newcommand{\FXY}{\ensuremath{f(x, y)}}

\newcommand{\XYZ}{\ensuremath{(x, y, z)}}

\newcommand{\ResolutionKrFullFourSevenThreeFour}{\SI{4.55 +- 0.01}{\percent}}
\newcommand{\ResolutionKrFullFourSevenThreeFourQbb}{\SI{0.592 +- 0.001}{\percent}}
\newcommand{\ResolutionKrFidFourSevenThreeFour}{\SI{3.88 +- 0.04}{\percent}}
\newcommand{\ResolutionKrFidFourSevenThreeFourQbb}{\SI{0.504 +- 0.005}{\percent}}

\newcommand{\ResolutionKrFullFourEightFourOne}{\SI{4.86 +- 0.01}{\percent}}
\newcommand{\ResolutionKrFullFourEightFourOneQbb}{\SI{0.631 +- 0.002}{\percent}}
\newcommand{\ResolutionKrFidFourEightFourOne}{\SI{3.93 +- 0.03}{\percent}}
\newcommand{\ResolutionKrFidFourEightFourOneQbb}{\SI{0.510 +- 0.004}{\percent}}

\newcommand{\ResolutionKrFullFourSevenThreeFourWithSystematics   }{\SI[parse-numbers=false]{(4.553  \pm 0.010 \ \stat \pm 0.324 \ \sys)}{\percent}}
\newcommand{\ResolutionKrFullFourSevenThreeFourQbbWithSystematics}{\SI[parse-numbers=false]{(0.5916 \pm 0.0014\ \stat \pm 0.0421\ \sys)}{\percent}}
\newcommand{\ResolutionKrFidFourSevenThreeFourWithSystematics    }{\SI[parse-numbers=false]{(3.804  \pm 0.013 \ \stat \pm 0.112 \ \sys)}{\percent}}
\newcommand{\ResolutionKrFidFourSevenThreeFourQbbWithSystematics }{\SI[parse-numbers=false]{(0.4943 \pm 0.0017\ \stat \pm 0.0146\ \sys)}{\percent}}

\newcommand{\ResolutionKrFullFourEightFourOneWithSystematics   }{\SI[parse-numbers=false]{(4.860  \pm 0.013 \ \stat \pm 0.246 \ \sys)}{\percent}}
\newcommand{\ResolutionKrFullFourEightFourOneQbbWithSystematics}{\SI[parse-numbers=false]{(0.6314 \pm 0.0017\ \stat \pm 0.0320\ \sys)}{\percent}}

\newcommand{\KrEnergy}{\SI{41.5}{\keV}}

\newcommand{\RbLifetime}{\SI{86.2}{\days}}
\newcommand{\RbIntensity}{\SI{1}{\kilo\becquerel}}
\newcommand{\KrLifetime}{\SI{1.83}{\hour}}
\newcommand{\KrLifetimeShort}{\SI{154.4}{\nano\second}}
\newcommand{\KrRate}{\SI{10}{\hertz}}
\newcommand{\KrPmtSumEminRunII}{\SI{5e+3}{\pes}}
\newcommand{\KrPmtSumEmaxRunII}{\SI{15e+3}{\pes}}

\newcommand{\KrSearchWindow}{\SI{620}{\micro\second}}
\newcommand{\KrSiPmThreshold}{\SI{10}{\pes}}
\newcommand{\KrTotVolumeRRunII}{\SI{200}{\mm}}
\newcommand{\KrFidVolumeRRunII}{\SI{150}{\mm}}
\newcommand{\KrExtFidVolumeRRunII}{\SI{175}{\mm}}
\newcommand{\KrFidVolumeZRunII}{\SI{150}{\mm}}

\newcommand{\KryptonLifetimeMapBinsRunII}{\ensuremath{60 \times 60}}
\newcommand{\KryptonEnergyMapPositionCraterRunII}{\ensuremath{[-50, 50]}}
\newcommand{\KryptonLifetimeMapBinSizeRunII}{\SI{6.7}{\mm}}

\newcommand{\KrXbMinusXt}{\SI{0.7}{\mm}}

\newcommand{\XenonIntrinsicEnergyResolution}{\SI{0.3}{\percent}}

\newcommand{\RunFourSevenThreeFourDate}{\DTMdisplaydate{2017}{10}{10}{-1}}

\newcommand{\RunFourSevenThreeFourTriggers}{\num{2687860}}

\newcommand{\RunFourSevenThreeFourTriggerRate}{\SI{10.5}{\hertz}}

\newcommand{\RunFourSevenThreeFourGateVoltage}{\SI{7.2}{\kV}}

\newcommand{\RunFourEightFourOneDate}{\DTMdisplaydate{2017}{11}{12}{-1}}
\newcommand{\RunFourEightFourOneTriggers}{\num{2993867}}
\newcommand{\RunFourEightFourOneTriggerRate}{\SI{8.2}{\hertz}}

\newcommand{\RunFourEightFourOneGateVoltage}{\SI{-8.5}{\kV}}

\newcommand{\NewGateVoltageSevenBarRunII}{\SI{-7.0}{\kV}}
\newcommand{\NewGateVoltageNineBarRunII}{\SI{-8.5}{\kV}}
\newcommand{\NewCathodeVoltageRunII}{\SI{-27}{\kV}}
\newcommand{\NewCathodeVoltageSevenBarRunII}{\SI{-28}{\kV}}
\newcommand{\NewCathodeVoltageNineBarRunII}{\SI{-30}{\kV}}

\newcommand{\NewGateVoltageRunII}{\SI{-7.2}{\kV}}

\newcommand{\NewCathodeVoltageAtFifteenBar}{\SI{-41}{\kV}}

\newcommand{\NewSevenBarPressureRunII}{\SI{7.2}{\bar}}
\newcommand{\NewNineBarPressureRunII}{\SI{9.1}{\bar}}

\newcommand{\NewIntrinsicEnergyResolution}{\SI{0.40}{\percent}}
\newcommand{\NewIntrinsicEnergyResolutionRunII}{\SI{0.45}{\percent}}
\newcommand{\NewPressureVesselMaterial}{316Ti}

\newcommand{\NewTpcLength}{\SI{664.5}{\mm}}

\newcommand{\NewTpcDriftLength}{\SI{530.3 +- 2}{\mm}}

\newcommand{\NewDriftField}{\SI{400}{\V\per\cm}}

\newcommand{\NewTpcELGap}{\SI{6}{\mm}}
\newcommand{\NewAnodePlateDiameter}{\SI{522}{\mm}}

\newcommand{\NewReducedField}{\SI{2.2}{\kV\per\cm\per\bar}}

\newcommand{\NewReducedFieldRunII}{\SI{1.7}{\kV\per\cm\per\bar}}

\newcommand{\NewGateVoltageAtFifteenBar}{\SI{-16.2}{\kV}}

\newcommand{\NewNumberOfSiPM}{\num{1792}}

\newcommand{\NewSipmPitch}{\SI{10}{\mm}}

\newcommand{\NewSiPMSampling}{\SI{1}{\micro\second}}

\newcommand{\NewNumberOfPMT}{\num{12}}

\newcommand{\NewCathodeToPMTs}{\SI{130}{\mm}}

\newcommand{\NewPMTSampling}{\SI{25}{\nano\second}}

\newcommand{\NewTpcDiameter}{\SI{454}{\mm}}
\newcommand{\NewFiducialMass}{\SI{5}{\kg}}
\newcommand{\NewFiducialMassSevenBar}{\SI{2.3}{\kg}}
\newcommand{\NewFiducialMassNineBar}{\SI{3}{\kg}}
\newcommand{\NewPressure}{\SI{15}{\bar}}

\newcommand{\NewTypePMT}{Hamamatsu R11410-10}
\newcommand{\NewPMTCoverage}{31\%}

\newcommand{\NewK}{0.016}

\newcommand{\NewBarrelICS}{\SI{60}{\mm}}

\newcommand{\NewPlatesICS}{\SI{120}{\mm}}

\newcommand{\NextTpcDiameter}{\SI{1050}{\mm}}
\newcommand{\NextTpcLength}{\SI{1300}{\mm}}

\begin{document}

\title{Calibration of the \NEW\ detector using \Kr{83m} decays}
% 
%\title{NEXT template for author-list and acknowledgments}

\collaboration{The NEXT Collaboration}
\author[17,18, a]{G.~Mart\'inez-Lema,\note[a]{Corresponding author.}}
\author[18]{J.A.~Hernando~Morata,}
\author[17]{B.~Palmeiro,}
\author[17]{A.~Botas,}
\author[14, 8]{P.~Ferrario,}
\author[14, 3]{F.~Monrabal,}
\author[17]{A.~Laing,}
\author[17]{J.~Renner,}
\author[17, 6]{A.~Sim\'on,}
\author[5]{A.~Para,}
\author[14, 8, b]{J.J.~G\'omez-Cadenas,\note[b]{NEXT Co-spokesperson.}}
\author[10]{C.~Adams,}
\author[17]{V.~\'Alvarez,}
\author[6]{L.~Arazi,}
\author[4]{C.D.R~Azevedo,}
\author[2]{K.~Bailey,}
\author[19]{F.~Ballester,}
\author[17]{J.M.~Benlloch-Rodr\'{i}guez,}
\author[12]{F.I.G.M.~Borges,}
\author[17]{S.~C\'arcel,}
\author[17]{J.V.~Carri\'on,}
\author[20]{S.~Cebri\'an,}
\author[12]{C.A.N.~Conde,}
\author[17]{J.~D\'iaz,}
\author[5]{M.~Diesburg,}
\author[12]{J.~Escada,}
\author[19]{R.~Esteve,}
\author[17]{R.~Felkai,}
\author[11]{A.F.M.~Fernandes,}
\author[11]{L.M.P.~Fernandes,}
\author[4]{A.L.~Ferreira,}
\author[11]{E.D.C.~Freitas,}
\author[14]{J.~Generowicz,}
\author[7]{A.~Goldschmidt,}
\author[18]{D.~Gonz\'alez-D\'iaz,}
\author[10]{R.~Guenette,}
\author[9]{R.M.~Guti\'errez,}
\author[2]{K.~Hafidi,}
\author[1]{J.~Hauptman,}
\author[11]{C.A.O.~Henriques,}
\author[9]{A.I.~Hernandez,}
\author[19]{V.~Herrero,}
\author[2]{S.~Johnston,}
\author[3]{B.J.P.~Jones,}
\author[17]{M.~Kekic,}
\author[16]{L.~Labarga,}
\author[5]{P.~Lebrun,}
\author[17]{N.~L\'opez-March,}
\author[9]{M.~Losada,}
\author[11]{R.D.P.~Mano,}
\author[10]{J.~Mart\'in-Albo,}
\author[17]{A.~Mart\'inez,}
\author[3]{A.~McDonald,}
\author[11]{C.M.B.~Monteiro,}
\author[19]{F.J.~Mora,}
\author[17]{J.~Mu\~noz Vidal,}
\author[17]{M.~Musti,}
\author[17]{M.~Nebot-Guinot,}
\author[17]{P.~Novella,}
\author[3,c]{D.R.~Nygren,\note[c]{NEXT Co-spokesperson.}}

\author[17,d]{J.~P\'erez,\note[d]{Now at Laboratorio Subterr\'aneo de Canfranc, Spain.}}
\author[17]{M.~Querol,}
\author[2]{J.~Repond,}
\author[2]{S.~Riordan,}
\author[15]{L.~Ripoll,}
\author[17]{J.~Rodr\'iguez,}
\author[3]{L.~Rogers,}
\author[17]{C.~Romo-Luque,}
\author[12]{F.P.~Santos,}
\author[11]{J.M.F. dos~Santos,}
\author[13,e]{C.~Sofka,\note[e]{Now at University of Texas at Austin, USA.}}
\author[17]{M.~Sorel,}
\author[13]{T.~Stiegler,}
\author[19]{J.F.~Toledo,}
\author[17]{J.~Torrent,}
\author[4]{J.F.C.A.~Veloso,}
\author[13]{R.~Webb,}
\author[13,f]{J.T.~White,\note[f]{Deceased.}}
\author[17]{N.~Yahlali}
\emailAdd{gonzalo.martinez.lema@usc.es}
\affiliation[1]{
Department of Physics and Astronomy, Iowa State University, 12 Physics Hall, Ames, IA 50011-3160, USA}
\affiliation[2]{
Argonne National Laboratory, Argonne, IL 60439, USA}
\affiliation[3]{
Department of Physics, University of Texas at Arlington, Arlington, TX 76019, USA}
\affiliation[4]{
Institute of Nanostructures, Nanomodelling and Nanofabrication (i3N), Universidade de Aveiro, Campus de Santiago, Aveiro, 3810-193, Portugal}
\affiliation[5]{
Fermi National Accelerator Laboratory, Batavia, IL 60510, USA}
\affiliation[6]{
Nuclear Engineering Unit, Faculty of Engineering Sciences, Ben-Gurion University of the Negev, P.O.B. 653, Beer-Sheva, 8410501, Israel}
\affiliation[7]{
Lawrence Berkeley National Laboratory (LBNL), 1 Cyclotron Road, Berkeley, CA 94720, USA}
\affiliation[8]{
IKERBASQUE, Basque Foundation for Science, Bilbao, E-48013, Spain}
\affiliation[9]{
Centro de Investigaci\'on en Ciencias B\'asicas y Aplicadas, Universidad Antonio Nari\~no, Sede Circunvalar, Carretera 3 Este No.\ 47 A-15, Bogot\'a, Colombia}
\affiliation[10]{
Department of Physics, Harvard University, Cambridge, MA 02138, USA}
\affiliation[11]{
LIBPhys, Physics Department, University of Coimbra, Rua Larga, Coimbra, 3004-516, Portugal}
\affiliation[12]{
LIP, Department of Physics, University of Coimbra, Coimbra, 3004-516, Portugal}
\affiliation[13]{
Department of Physics and Astronomy, Texas A\&M University, College Station, TX 77843-4242, USA}
\affiliation[14]{
Donostia International Physics Center (DIPC), Paseo Manuel Lardizabal, 4, Donostia-San Sebastian, E-20018, Spain}
\affiliation[15]{
Escola Polit\`ecnica Superior, Universitat de Girona, Av.~Montilivi, s/n, Girona, E-17071, Spain}
\affiliation[16]{
Departamento de F\'isica Te\'orica, Universidad Aut\'onoma de Madrid, Campus de Cantoblanco, Madrid, E-28049, Spain}
\affiliation[17]{
Instituto de F\'isica Corpuscular (IFIC), CSIC \& Universitat de Val\`encia, Calle Catedr\'atico Jos\'e Beltr\'an, 2, Paterna, E-46980, Spain}
\affiliation[18]{
Instituto Gallego de F\'isica de Altas Energ\'ias, Univ.\ de Santiago de Compostela, Campus sur, R\'ua Xos\'e Mar\'ia Su\'arez N\'u\~nez, s/n, Santiago de Compostela, E-15782, Spain}
\affiliation[19]{
Instituto de Instrumentaci\'on para Imagen Molecular (I3M), Centro Mixto CSIC - Universitat Polit\`ecnica de Val\`encia, Camino de Vera s/n, Valencia, E-46022, Spain}
\affiliation[20]{
Laboratorio de F\'isica Nuclear y Astropart\'iculas, Universidad de Zaragoza, Calle Pedro Cerbuna, 12, Zaragoza, E-50009, Spain}
\abstract{The \NEW\ (NEW) detector is currently the largest radio-pure high-pressure xenon gas time projection chamber with electroluminescent readout in the world. It has been operating at Laboratorio Subterr\'aneo de Canfranc (LSC) since October 2016. This paper describes the calibrations performed using \Kr{83m} decays during a long run taken from March to November 2017 (\RII). Krypton calibrations are used to correct for the finite drift-electron lifetime as well as for the dependence of the measured energy on the event transverse position which is caused by variations in solid angle coverage both for direct and reflected light and edge effects. After producing calibration maps to correct for both effects we measure an excellent energy resolution for 41.5 keV point-like deposits of \ResolutionKrFullFourSevenThreeFourWithSystematics\ FWHM in the full chamber and \ResolutionKrFidFourSevenThreeFourWithSystematics\ FWHM in a restricted fiducial volume. Using naive \SQRE\ scaling, these values translate into resolutions of \ResolutionKrFullFourSevenThreeFourQbbWithSystematics\ FWHM and \ResolutionKrFidFourSevenThreeFourQbbWithSystematics\ FWHM at the \Qbb\ energy of xenon double beta decay (\XenonQbb), well within range of our target value of 1\%.}

\arxivnumber{1804.01780}
\collaboration{\includegraphics[height=9mm]{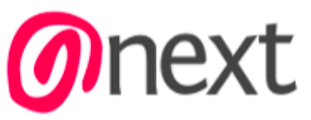}\\[6pt]
	NEXT collaboration}
%\collaboration{
%	NEXT collaboration}

\maketitle
\flushbottom

\section{\label{sec:introduction}Introduction}

The NEXT program is developing the technology of high-pressure xenon gas Time Projection Chambers (TPCs) with electroluminescent amplification (\HPXeEL) for neutrinoless double beta decay searches \cite{Nygren:2009zz, Alvarez:2011my, Alvarez:2012haa, Gomez-Cadenas:2013lta, Martin-Albo:2015rhw}. 
The first phase of the program included the construction, commissioning and operation of two prototypes, called NEXT-DEMO and 
NEXT-DBDM, which demonstrated the robustness of the technology, its excellent energy resolution and its unique topological signal \cite{Alvarez:2012xda, Alvarez:2013gxa, Alvarez:2012hh, Ferrario:2015kta}.
 
The \NEW\footnote{Named after Prof.~James White, our late mentor and friend.} (NEW) detector implements the second phase of the program. NEW is a $\sim$ 1:2 scale proof of principle detector for NEXT-100. The TPC has a length of \NewTpcLength\ and a diameter of \NewAnodePlateDiameter, while in the case of NEXT-100 the TPC has a length of \NextTpcLength\ and a diameter of  \NextTpcDiameter. NEXT-100 constitutes the third phase of the program with 100~kg of xenon and is foreseen to start operations in 2019.

\NEW\ has been running successfully since October 2016 at Laboratorio Subterr\'aneo de Canfranc (LSC). Its purpose is to validate the \HPXeEL\ technology in a large-scale radiopure detector.
This validation is composed of three main tasks: to assess the robustness and reliability of the technological solutions;
to compare in detail the background model with data, particularly the contribution to the radioactive budget of the different components, and
to study the energy resolution and the background rejection power of the topological signature characteristic of a \HPXeEL\ TPC.
Furthermore, \NEW\ can provide a measurement of the two-neutrino double beta decay mode (\bbtnu).
A \bbtnu\ signal significance of 8 $\sigma$ is expected for 100 days of run time \cite{Lopez-March:2017qzh}.
\par

After a short engineering run (Run I) in November-December 2016, the detector was operated continuously between March and November 2017 (Run II). In order to assess the overall performance of the detector and the gas system while minimizing the effect of possible leaks as well as limiting the energy of potential sparks, the operational pressure in Run II was limited to \NewSevenBarPressureRunII\ during the first part of data taking. The drift field was kept at \NewDriftField\  and the reduced field in the EL region at \NewReducedFieldRunII, slightly below the expected nominal value of \NewReducedField. This translates into a gate voltage of \NewGateVoltageRunII\ and a cathode voltage of \NewCathodeVoltageRunII. Under those conditions, the chamber was extremely stable, with essentially no sparks recorded over the period.

During the second part of the run the pressure was raised to  \NewNineBarPressureRunII\ and the voltages were correspondingly adjusted to keep approximately the same drift voltage and reduced field as during the first part of the run.

%Electron lifetime was low at the beginning of the data taking, possibly due to small oxygen leaks\footnote{This led to adding an exhaustive helium leak check of the full system before filling the chamber with xenon, which has been implemented for the current data run (Run II)}, but was improved, by continuous gas circulation through the getters to an acceptable value of near \NewLifetimeMaxRunII\ towards the end of the run. 
\par

In this paper, we describe the calibration of the detector using a rubidium source (\Rb{83}) which provides a large sample of krypton (\Kr{83m}) decays, yielding a high statistics sample of \KrEnergy\ energy deposits. These point-like, evenly-distributed events permit the continuous monitoring, measurement and correction for the drift-electron lifetime, as well as the measurement and correction for the dependence of the measured signal on the transverse \XY\ coordinates. After correcting both effects, the energy resolution of the chamber for point-like energy deposits can be determined.  

The paper is organized as follows: section \ref{sec.hpxe} explains the principle of operation of a \HPXeEL\ TPC and its intrinsic energy resolution; section \ref{sec.new} presents a brief description of the \NEW\ detector; section \ref{sec:krcal} describes how krypton calibrations are used to produce lifetime and energy maps; data processing and event selection are described in section \ref{sec:sel}; lifetime maps in section \ref{sec:lifetime}; energy maps in section \ref{sec:geometrical_corrections}; in section \ref{sec:energy_resolution} the energy resolution measured with \NEW\ is presented; and finally, results are summarized in section \ref{sec:summary}.

\section{Principle of operation and intrinsic resolution of a  \HPXeEL\ TPC}
\label{sec.hpxe}

\begin{figure}[bhtp!]
\centering
\subfloat[\label{fig:yield15}][15 bar ]{\includegraphics[width=0.45\textwidth]{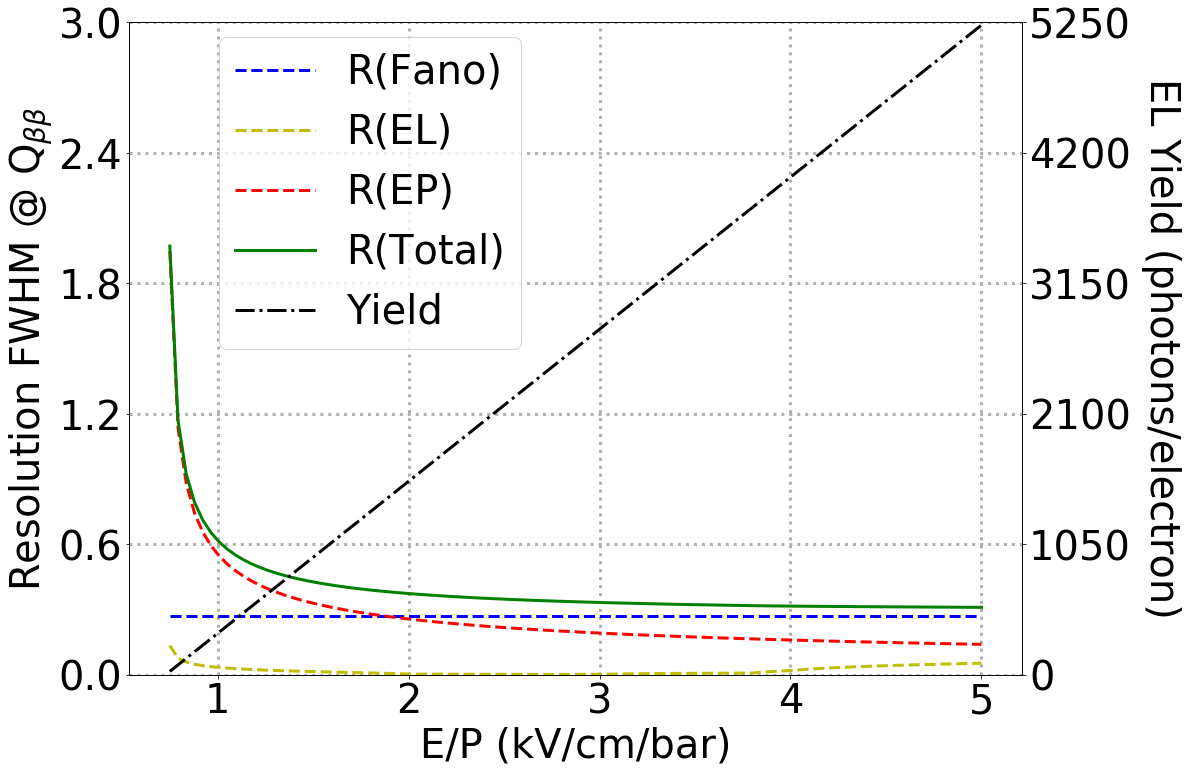}}
\hspace{4mm}
\subfloat[\label{fig:yield7 }][7.2 bar]{\includegraphics[width=0.45\textwidth]{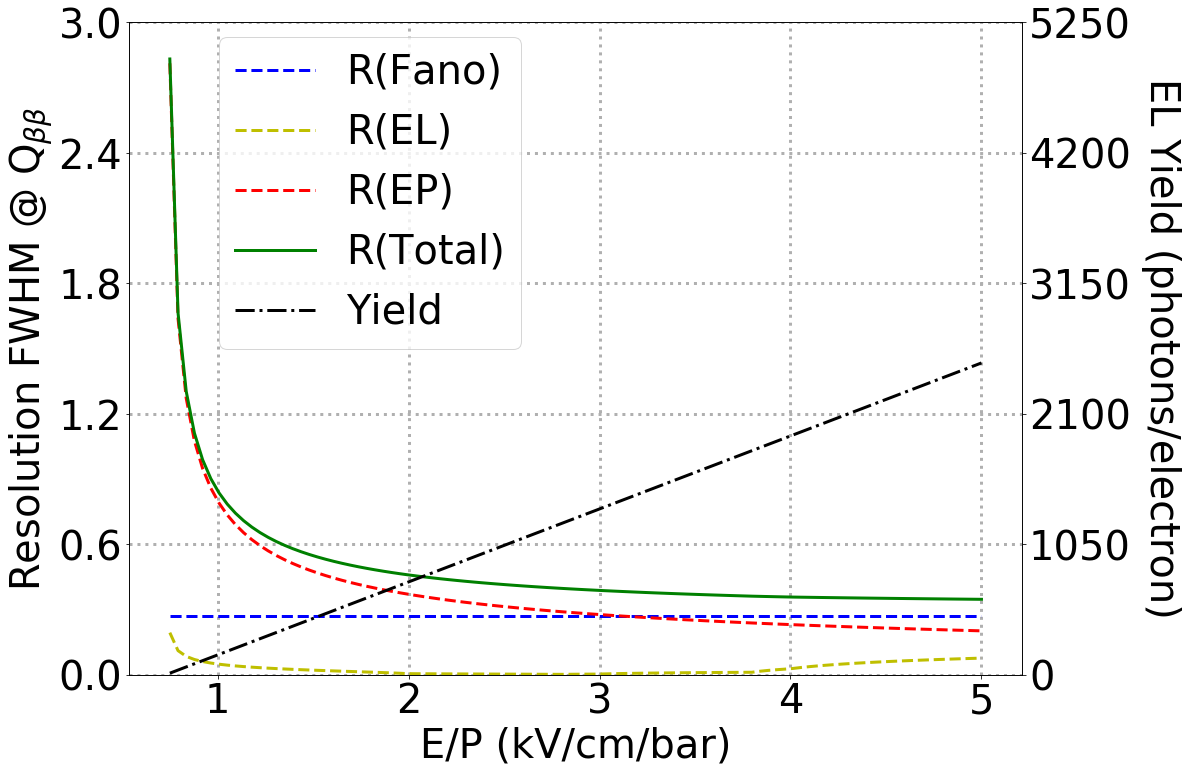}}
\caption{\small Energy resolution terms and EL yield characteristic of an \HPXeEL\ as a function of the reduced electric field for an EL gap of \NewTpcELGap\ and a value of $k \sim \NewK$ (see text for details). The left and right panels correspond, respectively, to pressures of \NewPressure\ and \NewSevenBarPressureRunII.}
\label{fig.yield}
\end{figure}
\begin{figure}[bhtp!]
\centering
\includegraphics[width=0.8\textwidth]{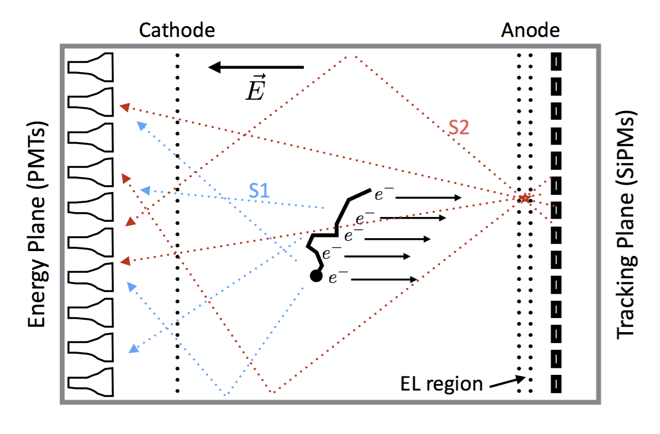}
\caption{\small Principle of operation of a \HPXeEL\ detector.}
\label{fig.po}
\end{figure}

As a charged particle propagates in the dense gas of a 
\HPXeEL TPC it loses energy by ionizing and exciting atoms of the medium. The excitation energy is manifested in the prompt emission of VUV (\XeWaveLength) scintillation light. The ionization electrons (ions) left behind by the particle are prevented from recombination and drifted towards the anode (cathode) by the action of a {\em drift field}. At the end of the field cage the drifting electrons enter the electroluminescent region (ELR), defined by a transparent mesh and a uniform layer of a titanium oxide compound (ITO) named the gate and the anode, respectively. The width of the ELR as well as the voltage drop between gate and anode are chosen to provide a reasonably large amplification of the ionization signal without further ionizing the gas. A  \HPXeEL\ is, therefore, an {\em optical TPC}, where both scintillation and ionization produce a light signal. In \NEW\ (and in NEXT-100), the light is detected by two independent sensor planes located behind the anode and the cathode. The energy of the event is measured by integrating the amplified EL signal (\st) with a plane of photomultipliers (PMTs). This {\em energy plane} also records the primary scintillation light (\so) which triggers the start-of-event (\tz).  
Electroluminescent light provides tracking as well, since it is detected a few mm away from production at the anode plane via a dense array of silicon photomultipliers (SiPMs), which constitute the  \emph{tracking plane}. This principle of operation is illustrated in \Fig\ \ref{fig.po}.

The reduced electroluminescent yield
(\ie, the number of photons produced per ionization electron entering the ELR, unit length and unit pressure),
$Y/P$, of an \HPXeEL\ can be empirically parametrized (\cite{Oliveira:2011xk}) as:
\begin{equation}
Y/P = (136 \pm 1) (E/P) - (99 \pm 4) {\rm ~[photons~electron^{-1}~cm^{-1}~bar^{-1}]}
\end{equation}
where $E/P$ is the reduced field in the ELR.

One of the desirable features of a \HPXeEL\ TPC is its excellent intrinsic energy resolution,
due to the small value of the Fano factor ($F$) \cite{PhysRev.72.26} in gaseous xenon and the small fluctuations of the EL yield.
Following \cite{Oliveira:2011xk}, the resolution (FWHM) of a \HPXeEL\ TPC can be written as:
\begin{equation}
 R_E = 2.35 \sqrt{ \frac{F}{\bar{N}_e}
 + \frac{1}{\bar{N}_e} (\frac{J}{\bar{N}_{EL}}) 
 + \frac{1 + (\sigma_q/q)^2}{\bar{N}_{ep}}}
 \label{eq.res}
\end{equation}
where
$\bar{N}_e$ is the number of primary ionization electrons created per event,
$\bar{N}_{EL}$ is the number of photons produced per ionization electron and $J$ its relative variance,
$\bar{N}_{ep}$ accounts for the total number of photoelectrons produced in the PMTs per event, and
the factor $\sigma_q/q$ represents the fluctuations in the photoelectron multiplication gain.
Thus, these terms represent, respectively, the fluctuations
in the number of primary electrons (a small number, given the low value of the Fano factor, $F=$ \XenonFanoFactor, in gaseous xenon),
in the electroluminescence yield and
in the PMT response (which in turn depends on the light collection efficiency of the detector and the distribution of the PMT single electron pulse heights).
For the \NEW\ PMTs $\sigma_q/q$ is approximately 0.5.
Moreover, since in pure xenon, $J / \bar{N}_{EL} \ll F$, equation \ref{eq.res} can be further simplified to \cite{Oliveira:2011xk}:
%$\bar{N}_{EL}$ is the number of photons produced per ionization electron and $J$ its relative variance, and
%$\bar{N}_{ep}$ accounts for the total number of photoelectrons produced in the PMTs per event.
%The factor $1.25$ takes into account both the conversion efficiency of VUV photons into photoelectrons, which provides a $\displaystyle\frac{1}{\bar{N}_{ep}}$ term, and the fluctuations in the photoelectron multiplication gain which provides an additional $\displaystyle\frac{1}{\bar{N}_{ep}} \cdot \left(\frac{\sigma_q}{\bar{G}_q}\right)^2$ term. The PMTs selected by NEXT have a relative charge resolution of $\sim$50\% which accounts for the additional $0.25$ in the numerator of the last term.
%Thus, these terms represent, respectively, the fluctuations
%in the number of primary electrons (a small number, given the low value of the Fano factor, $F=$ \XenonFanoFactor, in gaseous xenon),
%in the electroluminescence yield and
%in the PMT response (which in turn depends on the light collection efficiency of the detector and the distribution of the PMT single electron pulse heights).
%In pure xenon $J / \bar{N}_{EL} \ll F$, therefore, equation \ref{eq.res} can be further simplified to \cite{Oliveira:2011xk}:
%
\begin{equation}
 R_E = 2.35 \sqrt{\frac{F}{\bar{N}_e} + \frac{1.25}{\bar{N}_{ep}}}
 \label{eq.rest}
\end{equation}

In equation \ref{eq.rest} the Fano factor term gives the intrinsic energy resolution of xenon. 
For an energy of production of electron-ion pair of  
$w = \XenonEnergyPerElectron$ \cite{doi:10.1063/1.366105} and a value of the decay energy
$\Qbb = \XenonQbb$, one finds:
	
\begin{equation}
 \left. \bar{N}_e \right|_{\Qbb} =\frac{\Qbb}{w} = \XenonElectronsAtQbb
\end{equation}

and thus the intrinsic xenon resolution term has a constant value:

\begin{equation}
 \left. R_{int} \right|_{\Qbb} = 2.35 \sqrt{\frac{F}{\left. \bar{N}_e \right|_{\Qbb}}} \sim \XenonIntrinsicEnergyResolution
\end{equation}

The second term in equation \ref{eq.rest}, associated to the photon detection efficiency is inversely proportional to $N_{ep}$:
\begin{equation}
 \bar{N}_{ep} = k\ \bar{N}_e\ \bar{N}_{EL},
\end{equation}
where $k$~is the fraction of EL photons which go on to produce photoelectrons in the PMTs.

\Fig\ \ref{fig.yield} shows the resolution terms and the EL yield for a \HPXeEL\  TPC as a function of the reduced electric field for the following operational parameters: a) a pressure of \NewPressure\ (\NewSevenBarPressureRunII) corresponding to the nominal pressure of \NEXT\ (pressure of initial operation of \NEW);  b) an EL gap of \NewTpcELGap; and c) a value of $k \sim \NewK$ (corresponding to a light collection efficiency of \NewPMTCoverage). Notice that, while the resolution keeps improving with increasing reduced field, the improvement above \NewReducedField\ (a value for which the resolution reaches \NewIntrinsicEnergyResolution) is very small, while the required voltages at the gate become very large. For the initial operation of \NEW\ the reduced field was chosen to be \NewReducedFieldRunII, in order to limit the gate voltage to moderate values (\RunFourSevenThreeFourGateVoltage\ at a pressure of \NewSevenBarPressureRunII\ and \RunFourEightFourOneGateVoltage\ at a pressure of \NewNineBarPressureRunII). Under these operating conditions, the intrinsic resolution of the chamber at \Qbb\ (for point-like energy deposits) is \NewIntrinsicEnergyResolutionRunII. 

\section {Overview of the \NEW\ detector}
\label{sec.new}

\begin{table}[htp]
\caption{\NEW\ TPC parameters.}
\begin{center}
\begin{tabular}{|c|c|c|c|}
\hline
TPC parameter & Nominal & \RII\ (4734) & \RII\ (4841) \\
\hline
Pressure & \NewPressure & \NewSevenBarPressureRunII & \NewNineBarPressureRunII \\
EL field (E/P) & \NewReducedField & \NewReducedFieldRunII & \NewReducedFieldRunII \\
Drift field & \NewDriftField & \NewDriftField& \NewDriftField \\
$V_{cathode}$ &\NewCathodeVoltageAtFifteenBar &  \NewCathodeVoltageSevenBarRunII & \NewCathodeVoltageNineBarRunII\\
$V_{gate}$ & \NewGateVoltageAtFifteenBar &
                                           \NewGateVoltageSevenBarRunII & \NewGateVoltageNineBarRunII\\
\hline
Length & \NewTpcLength & \NewTpcLength & \NewTpcLength \\
Diameter &  \NewTpcDiameter & \NewTpcDiameter & \NewTpcDiameter \\
EL gap & \NewTpcELGap & \NewTpcELGap & \NewTpcELGap \\
Drift length & \NewTpcDriftLength & \NewTpcDriftLength & \NewTpcDriftLength \\
Fiducial mass & \NewFiducialMass & \NewFiducialMassSevenBar & \NewFiducialMassNineBar\\
\hline
\end{tabular}
\end{center}
\label{tab.TPC}
\end{table}%

The \NEW\ detector has been thoroughly described elsewhere \cite{Monrabal:2018xlr} and only a brief summary of its main features is offered here. It has three main subsystems, the TPC, the energy plane and the tracking plane. Table \ref{tab.TPC} shows the main parameters of the TPC. The
energy plane is instrumented with \NewNumberOfPMT\ \NewTypePMT\  PMTs located \NewCathodeToPMTs\ behind the cathode, providing a coverage of \NewPMTCoverage. The tracking plane is instrumented with \NewNumberOfSiPM\ SensL C-series SiPMs distributed in a square grid at a pitch of \NewSipmPitch. An ultra-pure \NewBarrelICS-thick copper shell (ICS) acts as a shield in the barrel region. The tracking plane and the energy plane are supported by \NewPlatesICS-thick pure copper plates.
 
The detector operates inside a pressure vessel made of stainless steel
\NewPressureVesselMaterial\ and is surrounded by a lead shield. Since a long electron lifetime is a must, xenon circulates in a gas system where it is continuously purified. The whole setup sits on top of a platform elevated over the ground in HALL-A of LSC. 

\section{Krypton calibrations}
\label{sec:krcal}

\begin{figure}[tbh!]
  \begin{center}
    \includegraphics[width=0.6\textwidth]{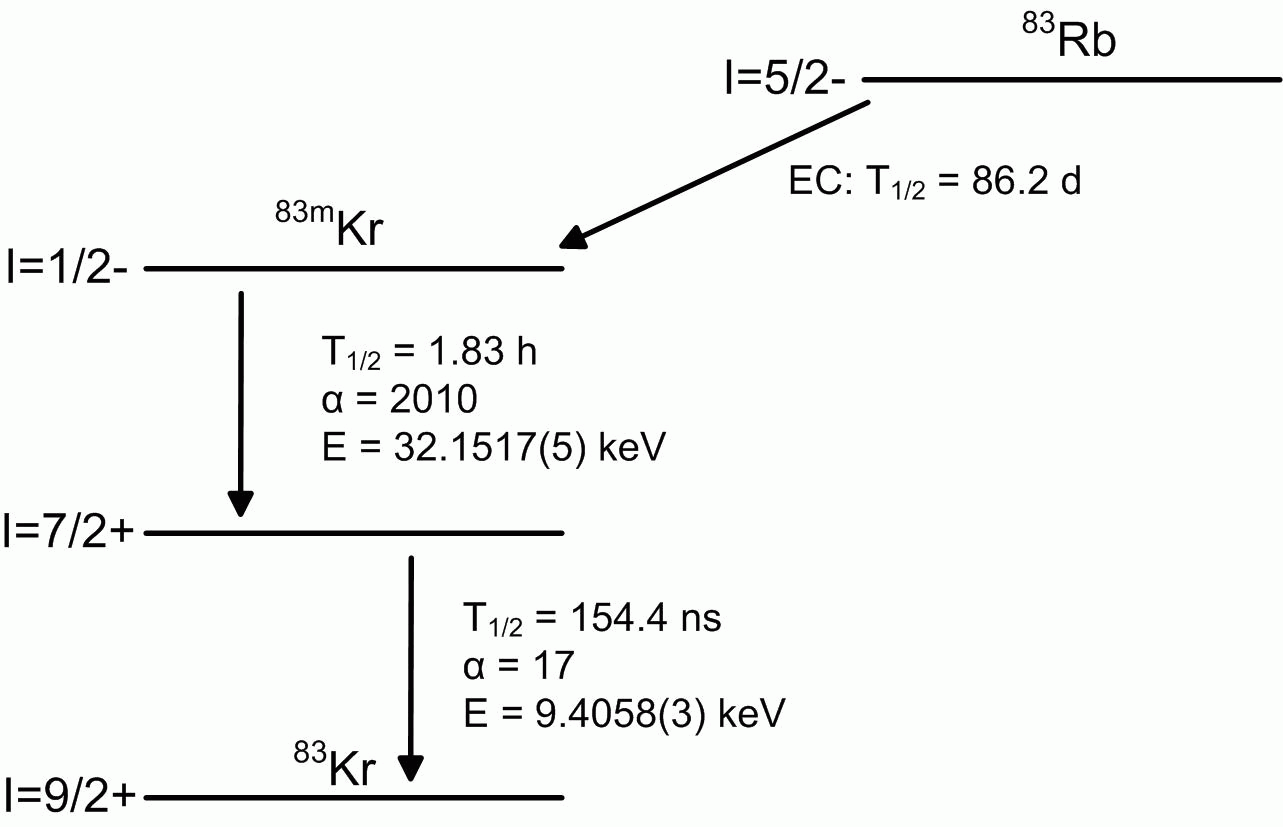} \hfill
    \caption{\label{fig:Rb_decay_scheme} \Rb{83} decay scheme.}
  \end{center}
\end{figure}

Figure \ref{fig:Rb_decay_scheme} shows the decay scheme of a \Rb{83} nucleus. The exotic rubidium isotope decays to \Kr{83m} via electron capture with a half-life of \RbLifetime. The krypton then decays to the ground state via two consecutive electron conversions. The decay rate is dominated by the first conversion with a half-life of \KrLifetime, while the second has a very short half-life of \KrLifetimeShort. The total released energy sums to \KrEnergy\ and the ground state of \Kr{83} is stable.

The rubidium source is formed by a number of small (1 mm-diameter) zeolite balls stored in a dedicated section of the gas system. \Kr{83m} nuclei produced after the electron capture of \Rb{83} emanate from the zeolite and flow with the gas inside the chamber. We do not expect a significant contribution to radon emanation from this section of the gas sytem (see \cite{Novella:2018ewv} for a detailed study of radon in \NEW). The \Rb{83} source has an intensity of \RbIntensity\ which results, after \Kr{83} decays in the gas system, in a rate of $\sim$100~Hz in the active volume. The rate of \Kr{83m} decay read-out is limited by the data acquisition system to a rate of about \KrRate\ to avoid DAQ crashes.

A \Kr{83m} decay results in a point-like energy deposition. The  time elapsed between detection of \so\ and detection of \st\ is the drift time and its measurement, together with the known value of the drift velocity \cite{Simon:2018vep}, determines the z-coordinate at which the ionization was produced in the active region. The \X\ and \Y\ coordinates are obtained by a position reconstruction algorithm which uses the charge recorded by the SiPMs of the tracking plane. The combination of the PMT and SiPM sensor responses yields a full 3D event reconstruction.

To properly measure the energy of an event in \NEW\, it is necessary to correct for two instrumental effects: a) {\em the finite electron lifetime}, due to attachment of ionization electrons drifting towards the cathode to residual impurities in the gas, and b) {\em the dependence of the light detected by the energy plane on the \XY\ position of the event}.  Krypton calibrations provide a powerful tool to measure and correct both effects. 

The effect of electron attachment is described using an exponential relation:

\begin{equation}
q(t) = q_0\ e^{-t/\LT}
\end{equation}
where $q_0$~ is the charge produced by the \Kr{83m} decay, $t$~ is the drift time, and \LT\ is the lifetime. Ideally, attachment can be corrected by measuring a single number. However, in a high pressure detector the lifetime may depend on the position \XYZ, due to the presence of non homogenous recirculation of the gas, or concentrations of impurities due to virtual leaks. As  discussed in section \ref{sec:lifetime}, the dependence of \LT\ with the longitudinal coordinate \Z\ in the \NEW\ detector can be neglected, while the dependence of \LT\ with the transverse \XY\ coordinates must be taken into account. This is done using krypton calibrations to produce a {\em lifetime map} that records the lifetime as a function of \XY.

Furthermore, \Kr{83m} decays can be used to produce a map of energy corrections. This map is needed to properly equalize the energy of events occurring in different locations in the chamber as the light detected by the photomultipliers depends on the \XY\ coordinates of the event even after \LT\ correction. Such dependence comes from the variation of the solid angle covered by the PMTs for direct light and expected acceptance for reflected light as well as from losses in events close to the detector edges. The map cannot be computed analytically due to the multiple reflections of the light inside the light tube and on other internal surfaces of the chamber.

%The calibration of detector response using meta stable krypton has been performed for a number of detectors with the general methodology being defined in the work described in  \cite{Kastens:2009pa}. The method was applied in liquid argon and neon detectors as described in \cite{PhysRevC.81.045803} and for liquid xenon by LUX as described in \cite{Akerib:2017eql}. Here, we report the first application of this type of calibration in a gaseous neutrino-less double beta decay detector.

\section{Data processing and event selection} 
\label{sec:sel}

\subsection*{Trigger}

\begin{figure}[tbh!]
  \begin{center}
    \includegraphics[width=0.49\textwidth]{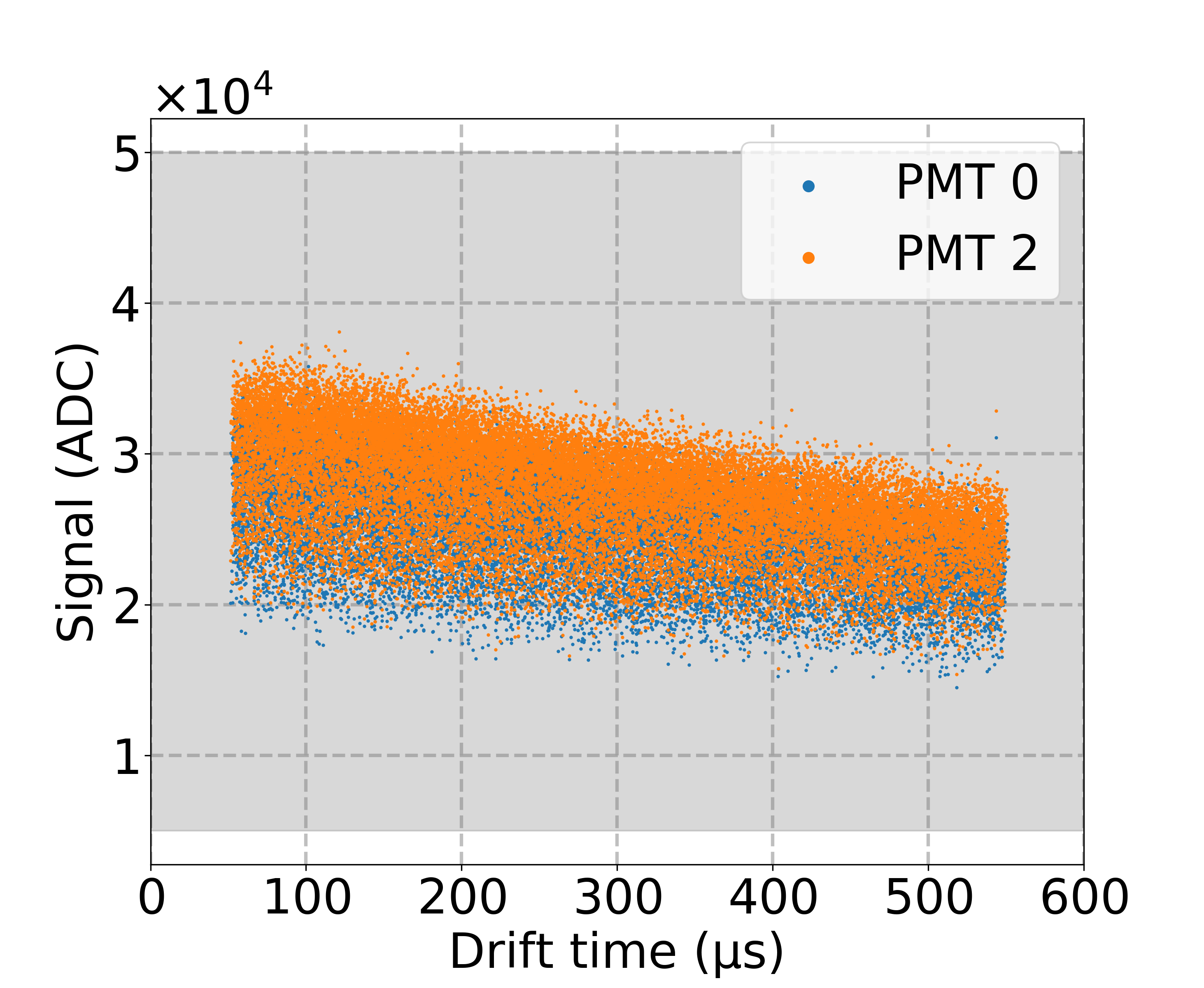}
    \includegraphics[width=0.49\textwidth]{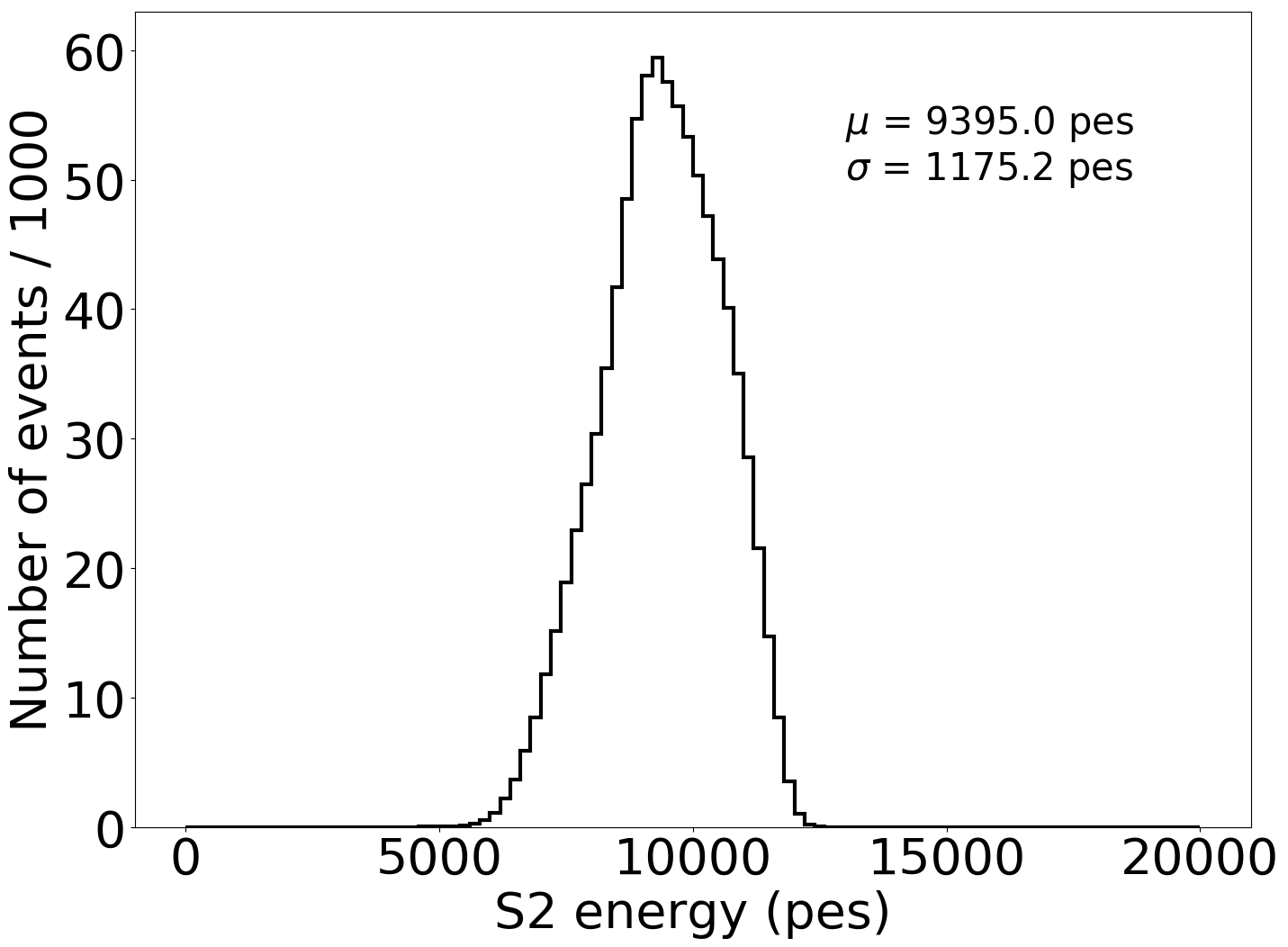}
    \caption{\label{fig:s2u} (left) Raw signal vs. drift for the two PMTs used in the trigger with a gray box indicating the limits imposed; (right) Uncorrected \st\ signal (in photoelectrons) measured by the sum of the PMTs.}
  \end{center}
\end{figure}

The detector triggers on the Krypton \st\ signals, specifically on the signal detected by two of the central PMTs. The trigger requires that these PMTs record a signal in the range 5000 and 50000 ADC (equivalent to approximately 180 to 1800 photoelectrons). The uncorrected  energy of the \st\ signals depends quite strongly on \XYZ\, given the sizable effect of both the spatial \XY\ corrections and the lifetime. However, as illustrated in \fig\ \ref{fig:s2u}, the trigger is open enough that no bias is expected and the range of the total signal detected in photoelectrons (pes) in the sum of the photomultipliers is well defined between \KrPmtSumEminRunII\ and \KrPmtSumEmaxRunII. %Each PMT accounts for roughly \num{1/12} of the total light recorded. For this reason the trigger requires that the two central PMTs record a signal in the range 200 to 1500 photoelectrons.

%The signal recorded in the 2 PMTs located at small radius is roughly 
%\num{1/12} of the total energy. Thus, events are triggered if the signal in {\em both} central PMTs is in the range between &\KrTriggerEminRunII\ and \KrTriggerEmaxRunII.

\subsection*{Waveform processing}

The raw data consists of PMT and SiPM waveforms. The PMT waveforms are sampled each \NewPMTSampling, while the SiPM waveforms are sampled each \NewSiPMSampling. The analysis proceeds according to the following steps.

\subsubsection*{\small Deconvolution of the PMT raw waveforms}

\begin{figure}[tbh!]
  \begin{center}
    \includegraphics[height=0.45\textwidth,width=0.49\textwidth, keepaspectratio]{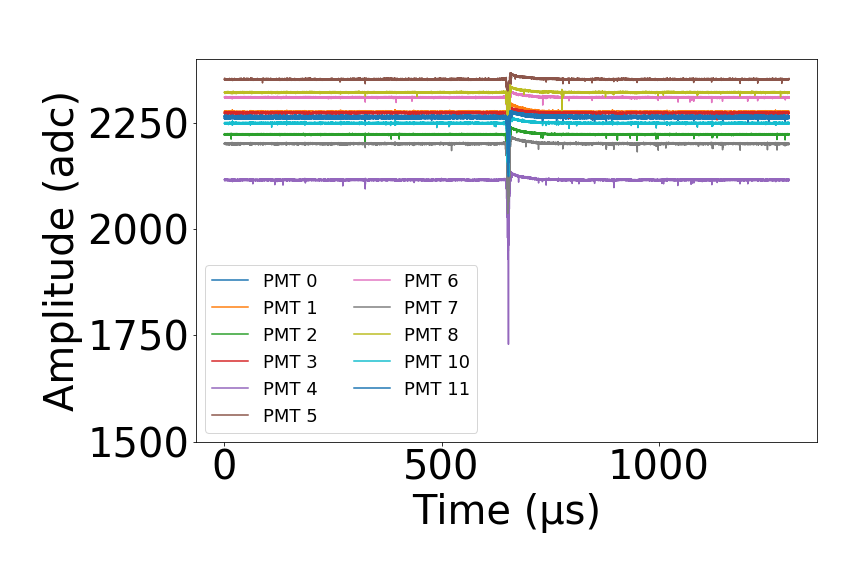}
    \includegraphics[height=0.45\textwidth,width=0.49\textwidth, keepaspectratio]{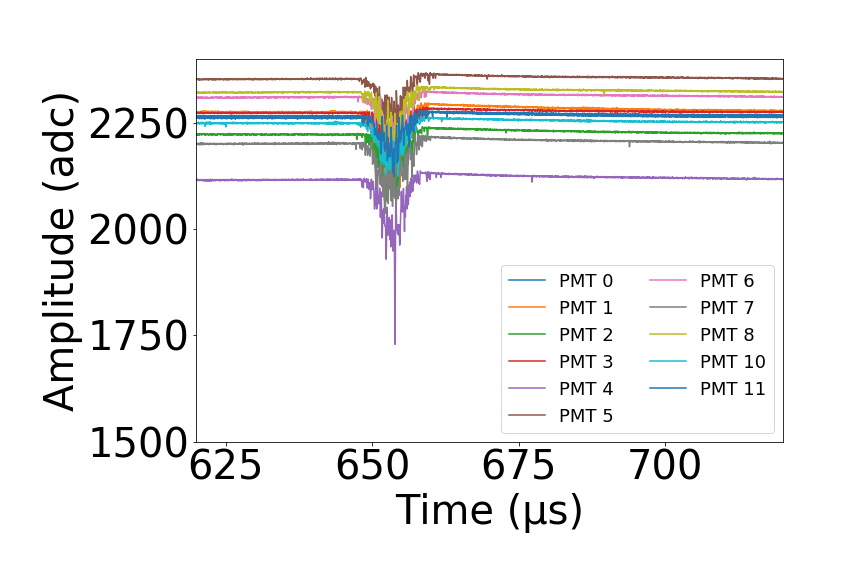}
    \caption{\label{fig:rwf} \Kr{83m} raw waveforms for the individual PMTs, showing the negative swing introduced by the PMT frond-end electronics. The left panel shows the RWF in the full DAQW, while the right panel shows a zoom on the \st\ signal on which the event was triggered.}
  \end{center}
\end{figure}

\begin{figure}[tbh!]
  \begin{center}
    \includegraphics[width=0.6 \textwidth]{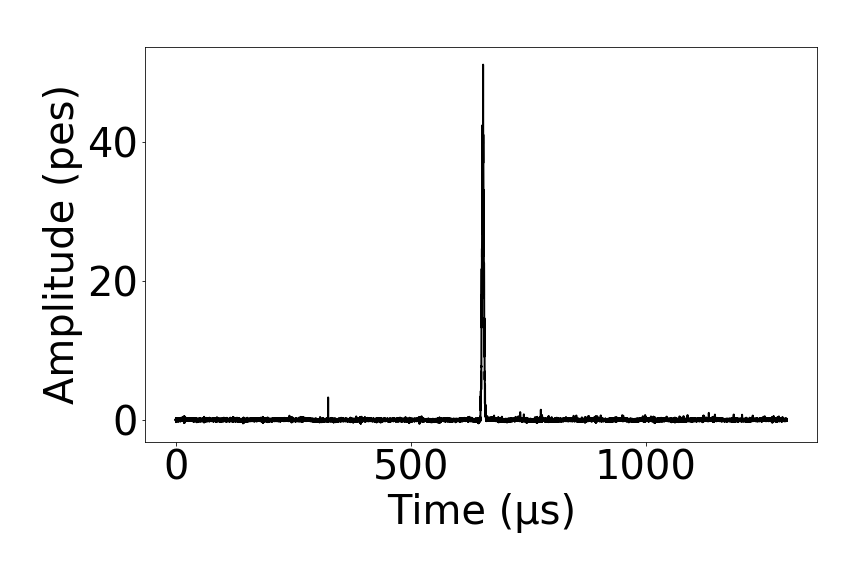}
    \includegraphics[width=0.45\textwidth]{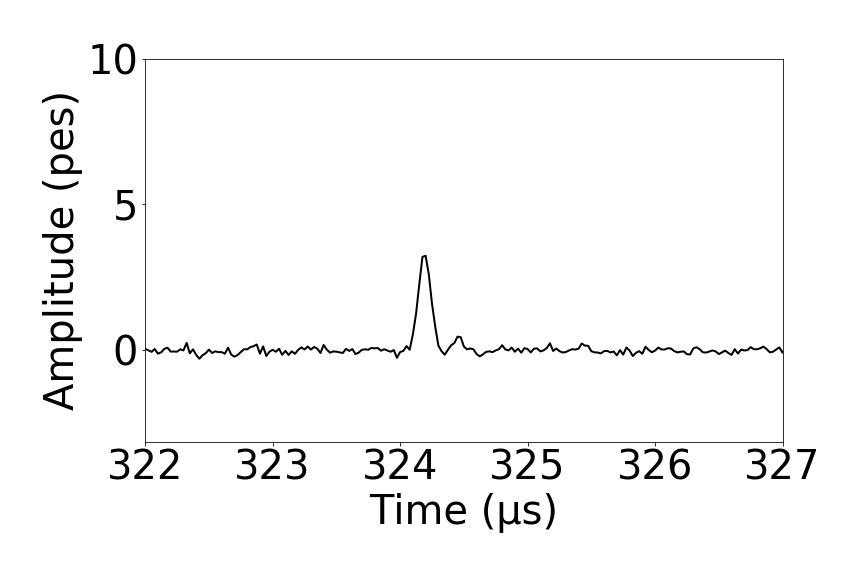}
    \includegraphics[width=0.45\textwidth]{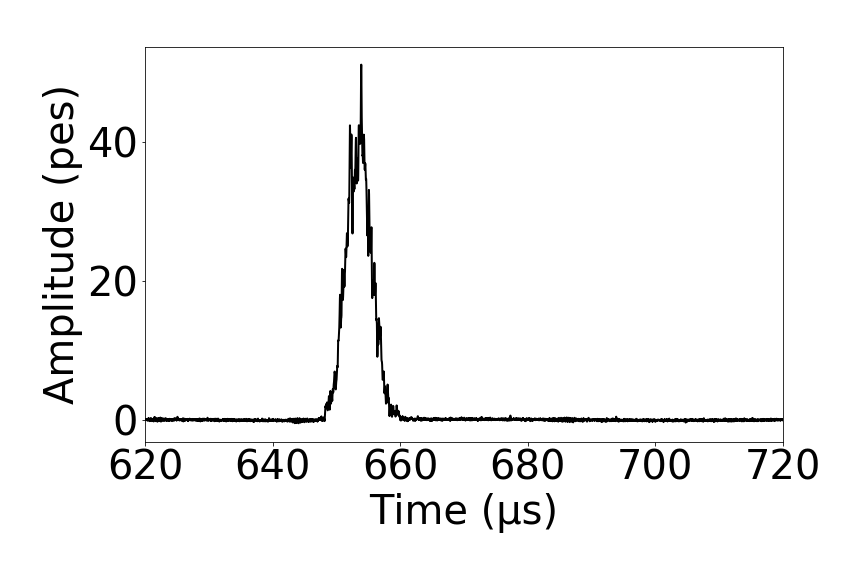}
    \caption{\label{fig:cwf} \Kr{83m} corrected waveforms for the sum of the PMTs. The top panel shows the CWF in the full DAQW, while the bottom panels show zooms of the \so\ (left) and \st\ (right) waveforms.}
  \end{center}
\end{figure}

As described in \cite{Monrabal:2018xlr}, the PMT waveforms show an opposite-sign swing due to the effect of the front-end electronics.
The first step in the processing is to apply a deconvolution algorithm \cite{Monrabal:2018xlr, Alvarez:2018xuc}, to the arbitrary-baseline, uncalibrated raw-waveforms (RWFs), to produce positive-only, zero-baseline, calibrated waveforms (CWFs).
\Fig\ \ref{fig:rwf} shows the RWFs corresponding to the PMTs of the energy plane, while  \Fig\ \ref{fig:cwf} shows the CWF waveform corresponding to the PMT sum.
The event was triggered by the \st\ signal, which appears centered in the data acquisition window (DAQW).
The \so\ signal appears up to the maximum drift time before the \st.

\subsubsection*{\small Search for signals}

\begin{figure}[tbh!]
  \begin{center}
    \includegraphics[width=0.5\textwidth]{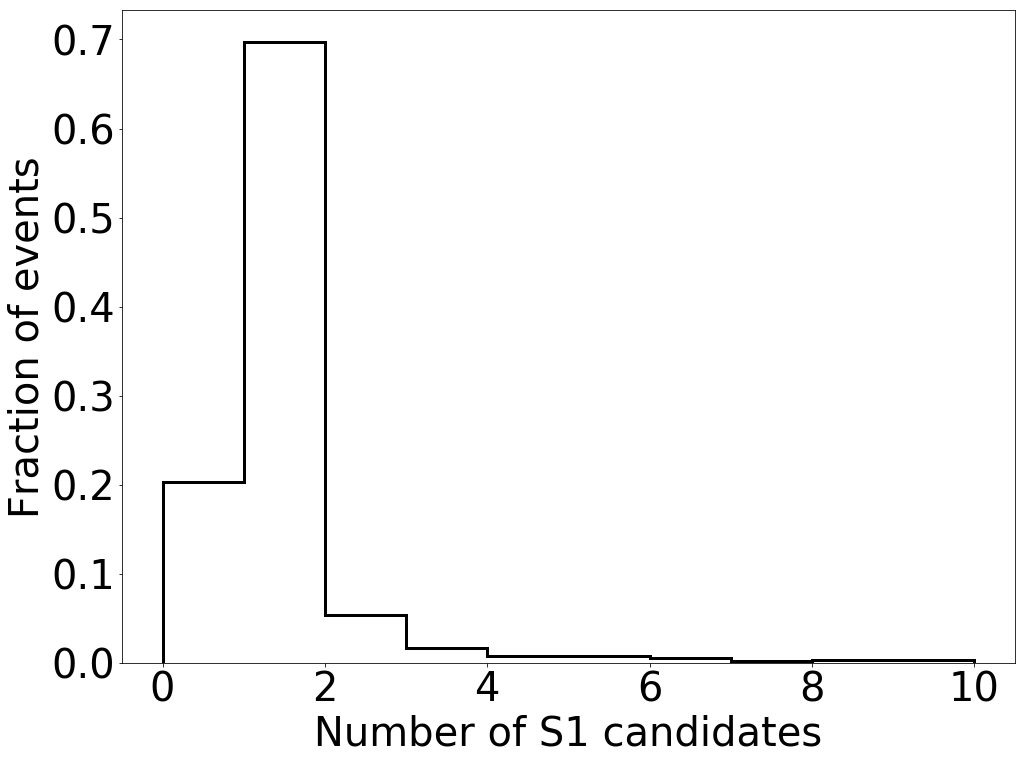}
    \caption{\label{fig:ns1} Number of \so\ candidates found by the peak-finding algorithm.}
  \end{center}
\end{figure}

The first half of the CWF sum (buffer time below \KrSearchWindow) is processed by a peak-finding algorithm tuned to find small (\so) signals.
The distribution of the number of \so\ candidates is shown in \fig\ \ref{fig:ns1}.
A single \so\ is identified in about 70 \% of the events, while two or more \so\ candidates are identified in near 10 \% of the events and 20 \% lack a \so\ signal.
The one-\so\ sample is dominated by genuine \so\ signals, while the sample with two or more \so\ include fake signals associated with krypton events happening in the field-cage buffer, or small scintillation signals.
Only events with exactly one \so\ candidate pass to the next stage of the selection.

The second half of the CWF sum (buffer time greater than \KrSearchWindow) is then processed by the same peak-finding algorithm, this time tuned to find larger signals. Most of the time a single \st\ candidate is found. Only events with exactly one \so\ and one \st\ are accepted for the analysis. 
 
\subsubsection*{\small Position of the event}

\begin{figure}[tbh!]
  \begin{center}
    \includegraphics[width=0.7\textwidth]{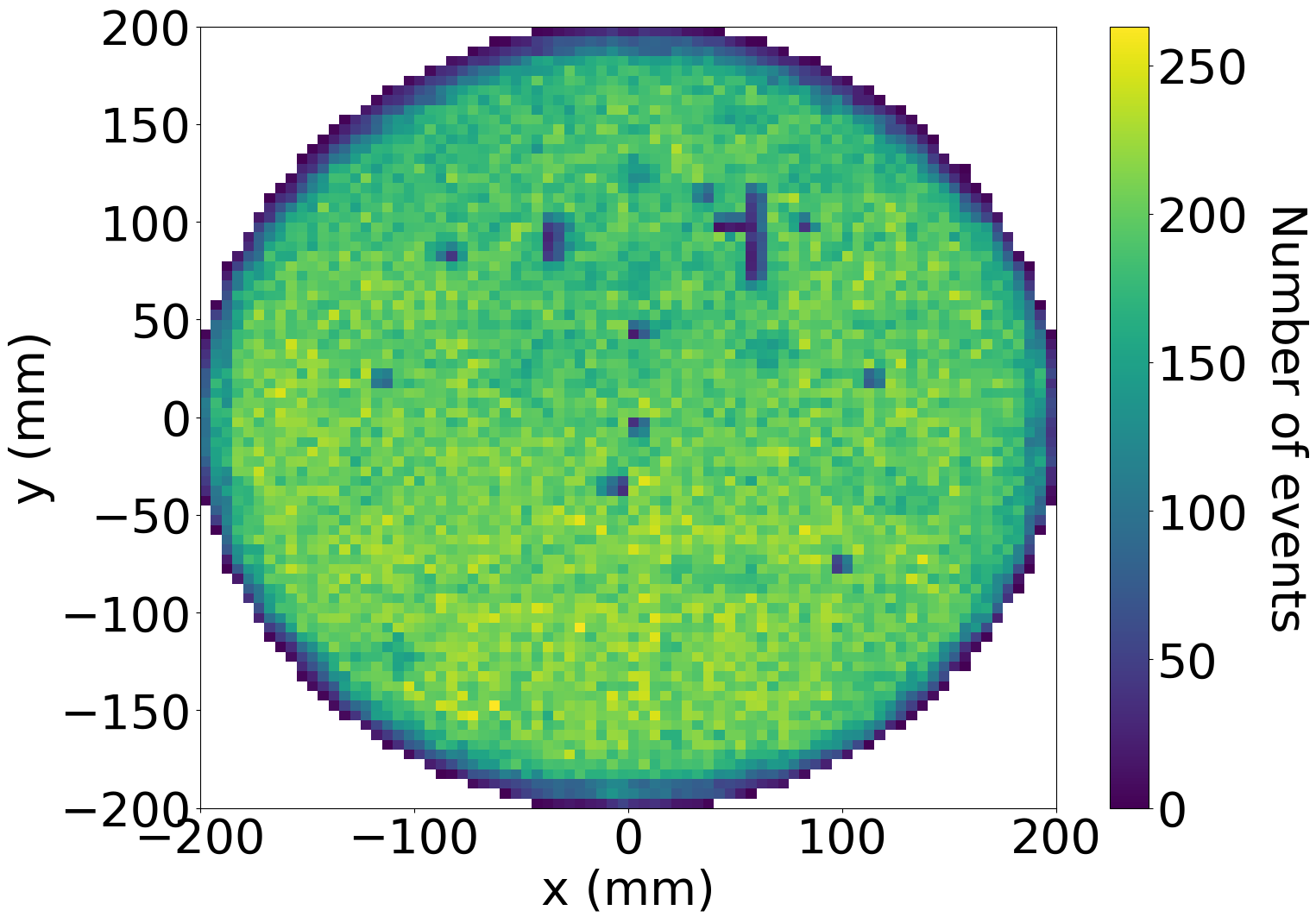} \hfill
    \caption{\label{fig:xy} Distribution of events in the \XY\ plane.}
  \end{center}
\end{figure}

\begin{figure}[tbh!]
  \begin{center}
    \includegraphics[width=0.7\textwidth]{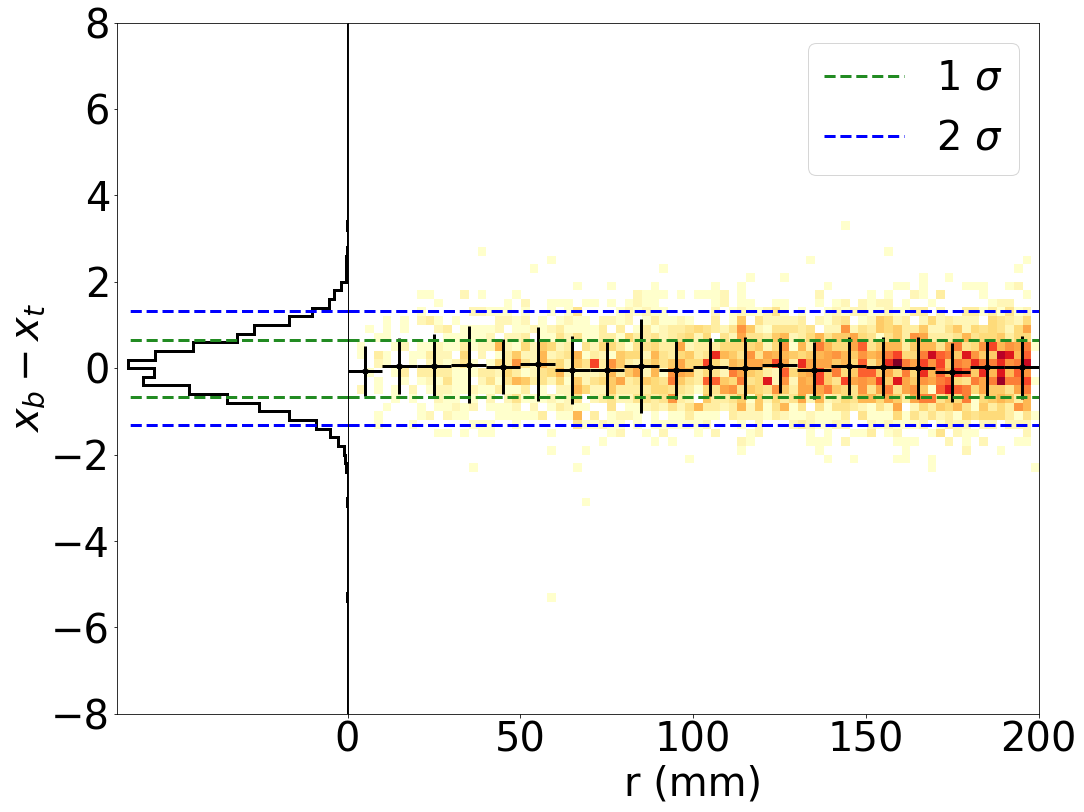}
    \caption{\label{fig:baricenter} Main panel: difference between the reconstructed and true \X\ position, $\Delta x$, for Monte Carlo krypton events as a function of the radial position; left sub-panel: distribution of the $\Delta x$ variable. The standard deviation of the distribution is (0.663 $\pm$ 0.010) mm. A similar distribution is found for the \Y\ coordinate.}
  \end{center}
\end{figure}

The \Z\ coordinate of the events is computed by multiplying the drift time (obtained as the difference
between \so\ and \st\ times) by the drift velocity, \VD, which is also measured using the data themselves as described in \cite{Simon:2018vep}. The position and charge of each SiPM with signal above \KrSiPmThreshold\ (chosen to eliminate spurious hits due to SiPM dark counts) within the waveform range defined by the \st\ are used to calculate a local barycentre around the SiPM with maximum signal which estimates the \XY\ position of the event. 

\Fig\ \ref{fig:xy} shows the distribution of events in the \XY\ plane, which is roughly uniform. The low-statistics pixels (dark blue color) correspond to inoperative or defective SiPMs or to defects in the ELR. \Fig\ \ref{fig:baricenter} shows that the reconstruction algorithm is well behaved.
The main panel displays the difference between the reconstructed and true \X\ position, $\Delta x$, as a function of the true radial coordinate of Monte Carlo krypton events.
The mean with $\pm$1 standard deviation, displayed as error bars, confirms that the algorithm is not biased at large radii.
The left sub-panel displays the distribution of $\Delta x$, forming a gaussian distribution with $\mu = 0$ and $\sigma = $\KrXbMinusXt.
A similar distribution is found for the \Y\ coordinate.

Monte Carlo events have been generated using a GEANT4-based program \cite{Agostinelli:2002hh} which incorporates a detailed description of the geometry and materials of the detector, the simulation of \Kr{83m} decays, the light propagation of \so\ and \st\ signals, and the response of the SiPMs and PMTs sensors.
A thorough description of the program is given in \cite{Martin-Albo:2015dza}.
The output of the program are simulated waveforms (similar to those CWF obtained from data). The same selection and reconstruction process applied to data have been applied to simulated events.

\subsubsection*{Datasets}

The data used in this analysis were collected with \NEW\ in Fall 2017. Two runs are considered. Run 4734 started on  \RunFourSevenThreeFourDate\ and collected \RunFourSevenThreeFourTriggers\ events at a trigger rate of \RunFourSevenThreeFourTriggerRate. The pressure was \NewSevenBarPressureRunII, the cathode was held at \NewCathodeVoltageSevenBarRunII\ and the gate at \NewGateVoltageSevenBarRunII. Run 4841 started on  \RunFourEightFourOneDate\ and collected \RunFourEightFourOneTriggers\ events at a trigger rate of \RunFourEightFourOneTriggerRate. The pressure was \NewNineBarPressureRunII, the cathode was held at \NewCathodeVoltageNineBarRunII\ and the gate at \NewGateVoltageNineBarRunII.

\section{Lifetime maps}
\label{sec:lifetime} 

As stated in section \ref{sec:krcal},  under certain conditions, (non homogenous recirculation of the gas combined with concentrations of impurities due to virtual leaks), the drift lifetime, \LT, may depend on the \XYZ\ position. 
\begin{figure}[tbh!]
  \begin{center}
    \includegraphics[width=0.65\textwidth]{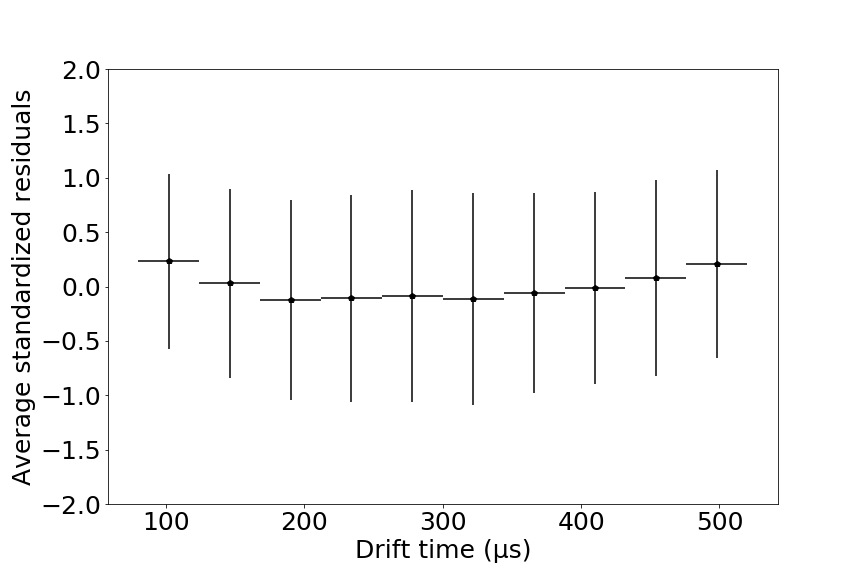}
    \caption{\label{fig:lres} Average electron-lifetime (\LT) residuals ($R_z$) as a function of
    \Z, showing that \LT\ is practically constant along the longitudinal coordinate and the rms of the pulls is stable at $\sim$1.}
  \end{center}
\end{figure}

In \NEW\ the dependence of \LT\ with \Z\ is found to be negligible. This is illustrated by 
\fig\ \ref{fig:lres}, where the dependence of the average residuals, 
$R_z = \displaystyle\frac{\sum_{xy} r_z}{n_{xy}}$~ is plotted as a function of
\Z. $R_z$ is computed by dividing the chamber into $n_{xy}$ bins in the transverse coordinates \XY\ and fitting the function $f(t) = e_0\ e^{-t/\LT}$ to the detected signal versus drift time (the raw \Z\ coordinate) in each \XY\ bin. The pull in each bin for each fit is then calculated as $r_z = \displaystyle\frac{e_t - f(t)}{\sigma}$. The distribution of these values over the \XY\ plane for each \Z\ bin is then fitted with a Gaussian and the mean value taken as $R_z$. As can be seen in \fig\ \ref{fig:lres}, the dependence of this value with \Z\ is very small, justifying the hypothesis that the \LT\ does not depend on \Z.
The data correspond to run 4734, but run 4841 shows the same behavior.
%To compute $R_z$~the chamber is divided first in $n_{xy}$~bins in the transverse coordinates \XY.  Each \XY\ bin is in turn divided in $n_z$~bins wich are used as points to fit an exponential parametrised as $f(t) = e_0\ e^{-t/\LT}$ for each \XY\ bin, where $e_0$ is the expected signal at zero drift. The residuals are then calculated as $r_z = \displaystyle\frac{e_t - f(t)}{\sigma}$. This residual quantifies the difference (normalized to the measurement uncertainty, $\sigma$) between the energy measured in the corresponding \Z\ bin ($e_t$) and the expected value of the energy in
%the bin ($f(t)$). %using a single \LT, \ie, $ f(z) = e_0\ e^{-t/(\LT v_d)}$, where $v_d$~is the drift velocity and $e_0$~the energy in the bin $z=0$.
%As can be seen, the dependence of $R_z$~with drift is very small, thus justifying the hypothesis that the \LT\ does not depend on \Z.
%The data correspond to run 4734, but run 4841 shows the same behavior. 

\begin{figure}[tbh!]
  \begin{center}
    \includegraphics[width=0.45\textwidth]{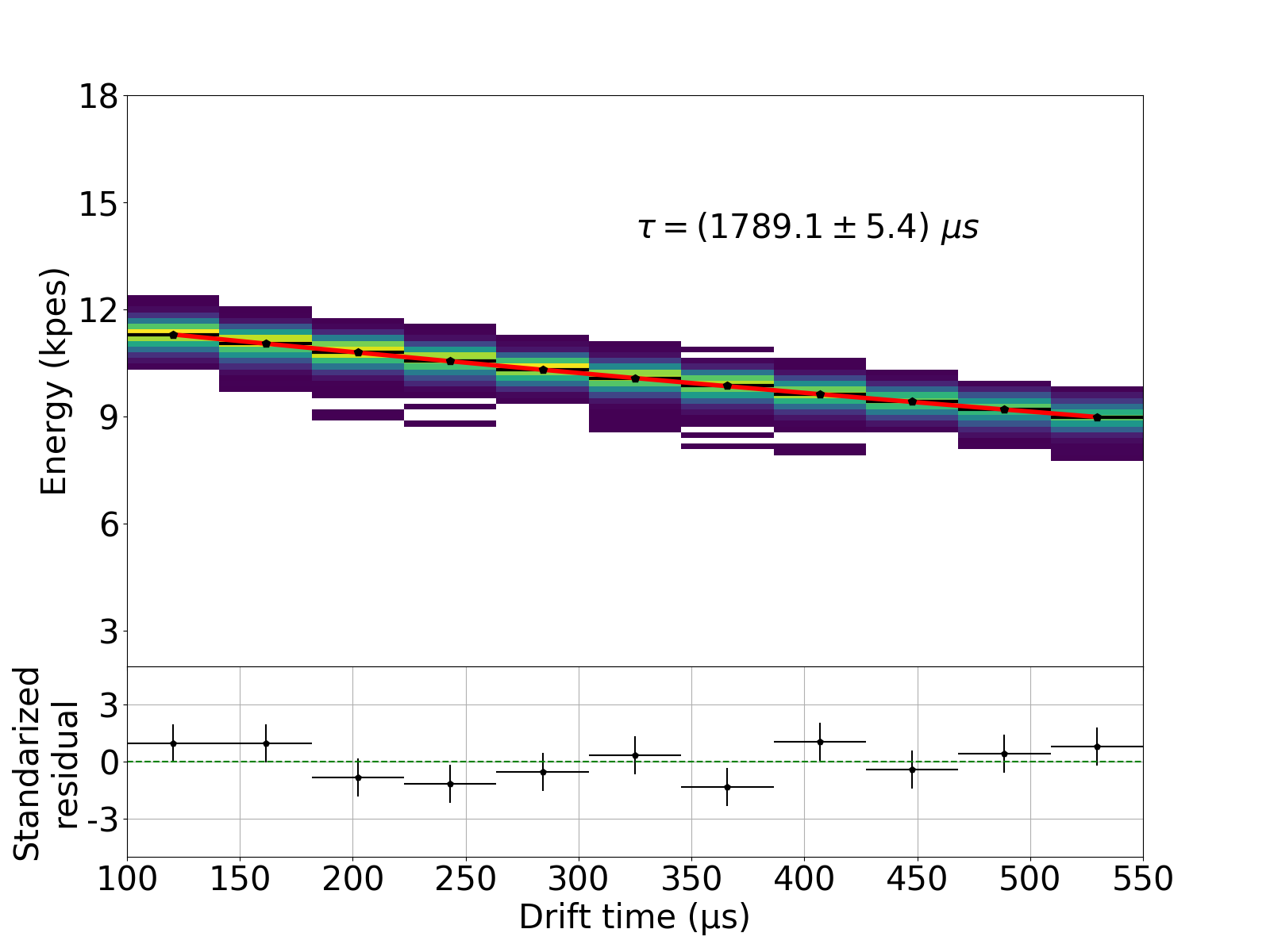}
    \hspace{5mm}
    \includegraphics[width=0.45\textwidth]{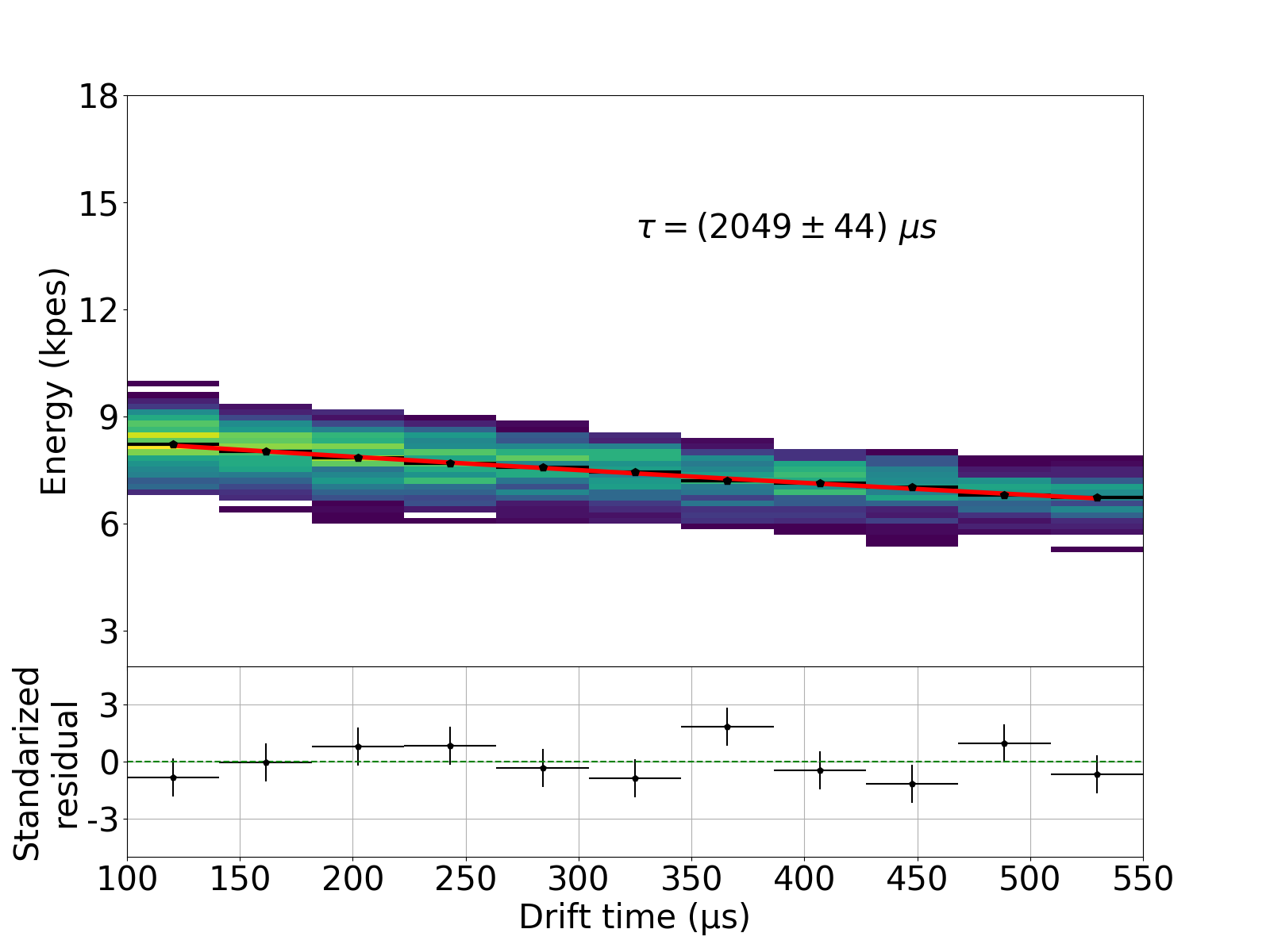}
    \caption{\label{fig:lifetime} Exponential fits to the distribution of krypton integrated signal as a function of the drift time in two different regions of the chamber.
             In the region defined by $\rm{x =  [0, 50]~mm}    $, $\rm{y = [0, 50]~mm}   $ (left panel) the lifetime is \SI{1789 +- 5}{\micro\second}, whilst
             in the region defined by $\rm{x =  [120,  150]~mm}$, $\rm{y = [120, 150]~mm}$ (right panel) the lifetime is \SI{2049 +- 44}{\micro\second}.
             Color indicates number of events, black lines averages and the red line the best fit line.}
  \end{center}
\end{figure}

On the other hand, the lifetime is found to depend on \XY\ for run 4734. This effect is illustrated in 
\fig\ \ref{fig:lifetime} where lifetime fits for two regions are shown, one near the center, defined by  $\rm{x = [0, 50]~mm}$, $\rm{y = [0, 50]~mm}$ (left panel) and one in the upper edge, defined by $\rm{x = [120,  150]~mm}$,\linebreak $\rm{y = [120, 150]~mm}$ (right panel). A reduction in statistics towards the cathode is seen in many \XY\ bins. The reduction is expected due to the combination of the decay time of the isotope and the complex flow of gas within the detector active region. The statistics are, however, sufficient that we find no bias in the determination of any calibration constants. All fits yield a good \chitwo$/$dof (0.9 for the first example and 1.0 for the second being typical values), but considerably different lifetimes of
\SI{1789 +- 5 }{\micro\second} (near the center of the chamber) and 
\SI{2049 +- 44}{\micro\second} (near the top of the chamber). 

\begin{figure}[tbh]
  \begin{center}
    \includegraphics[width=0.4 \textwidth]{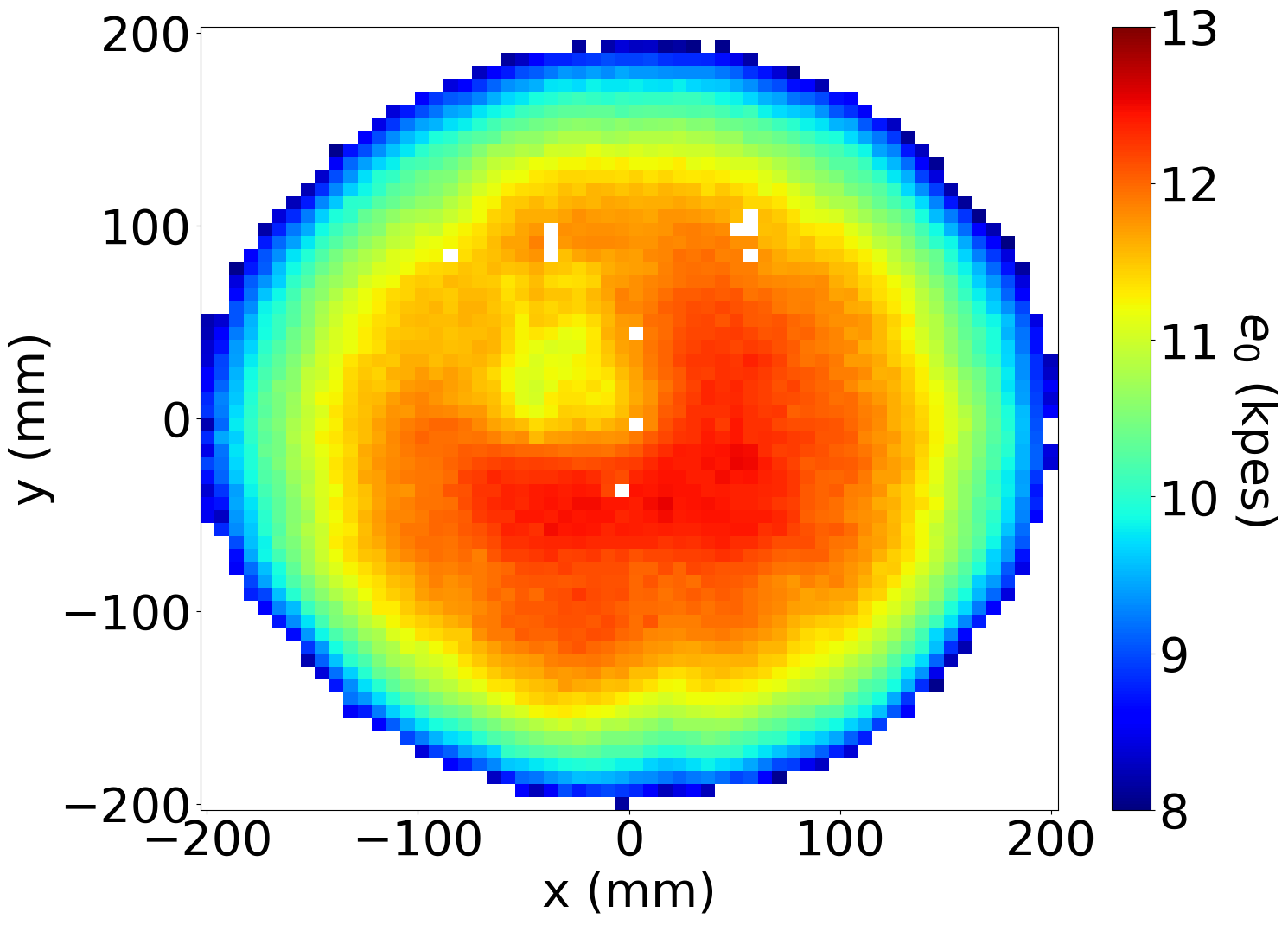}
    \hspace{5mm}
    \includegraphics[width=0.4 \textwidth]{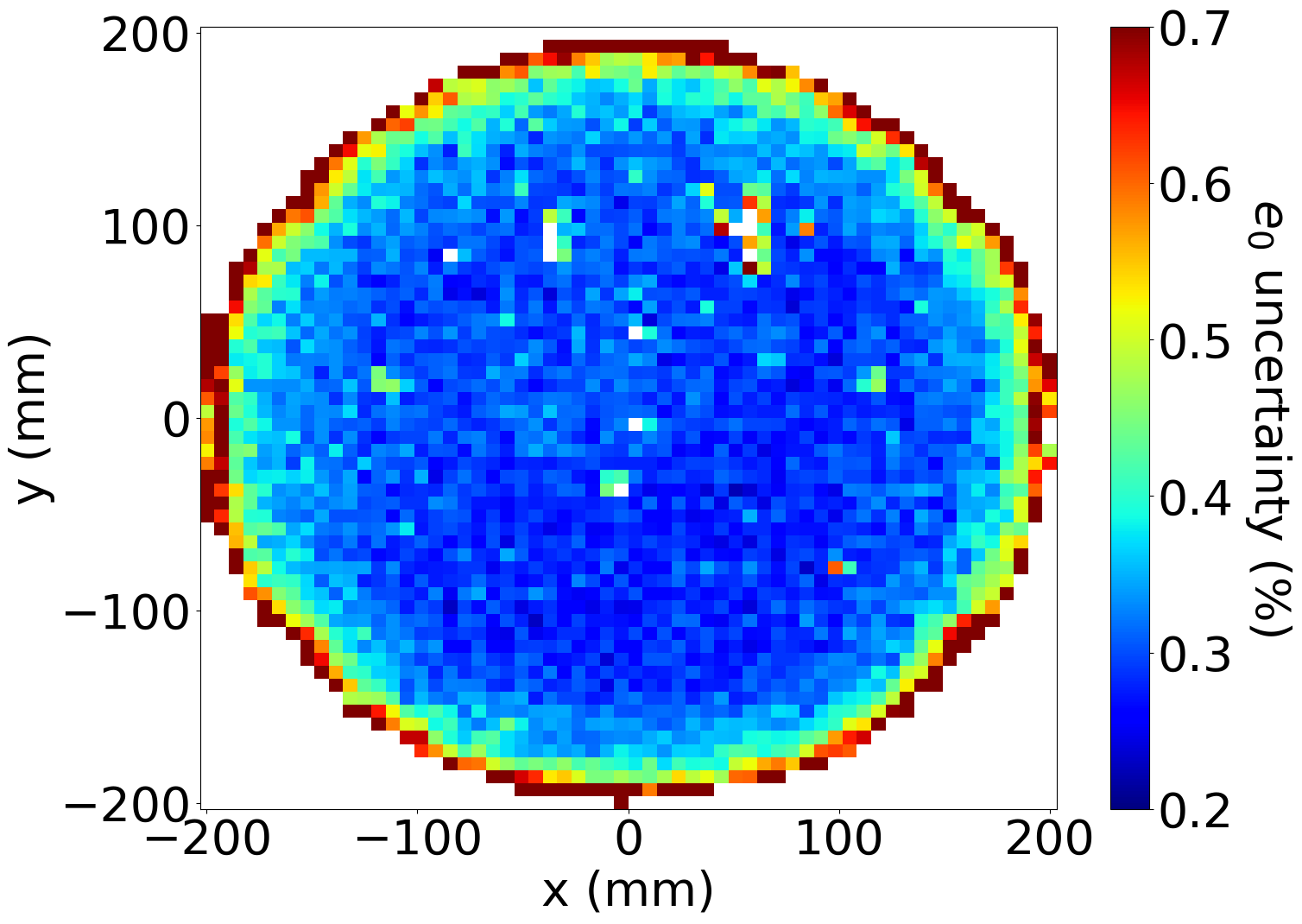}
    \vspace{5mm}
    \includegraphics[width=0.4 \textwidth]{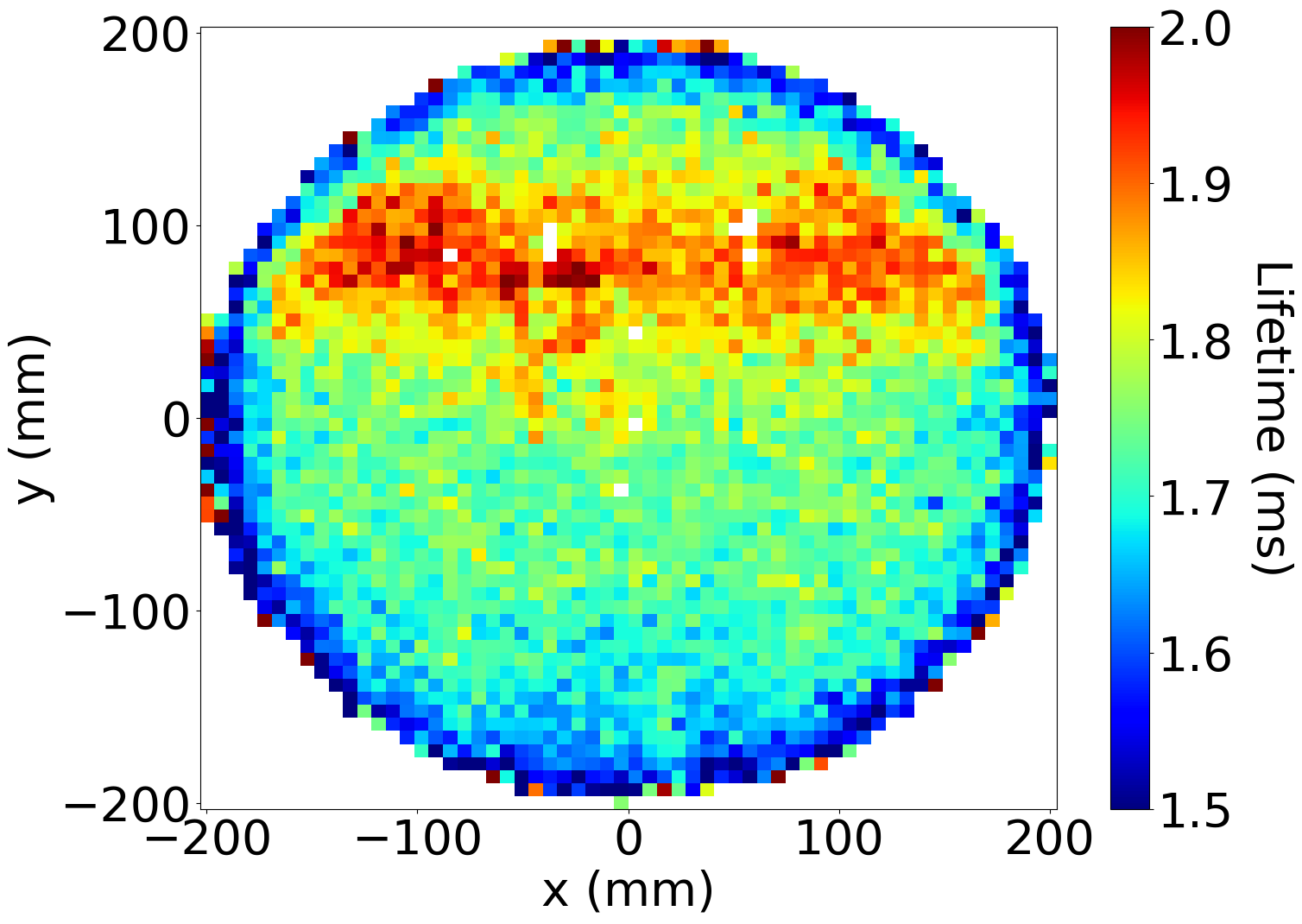}
    \hspace{5mm}
    \includegraphics[width=0.4 \textwidth]{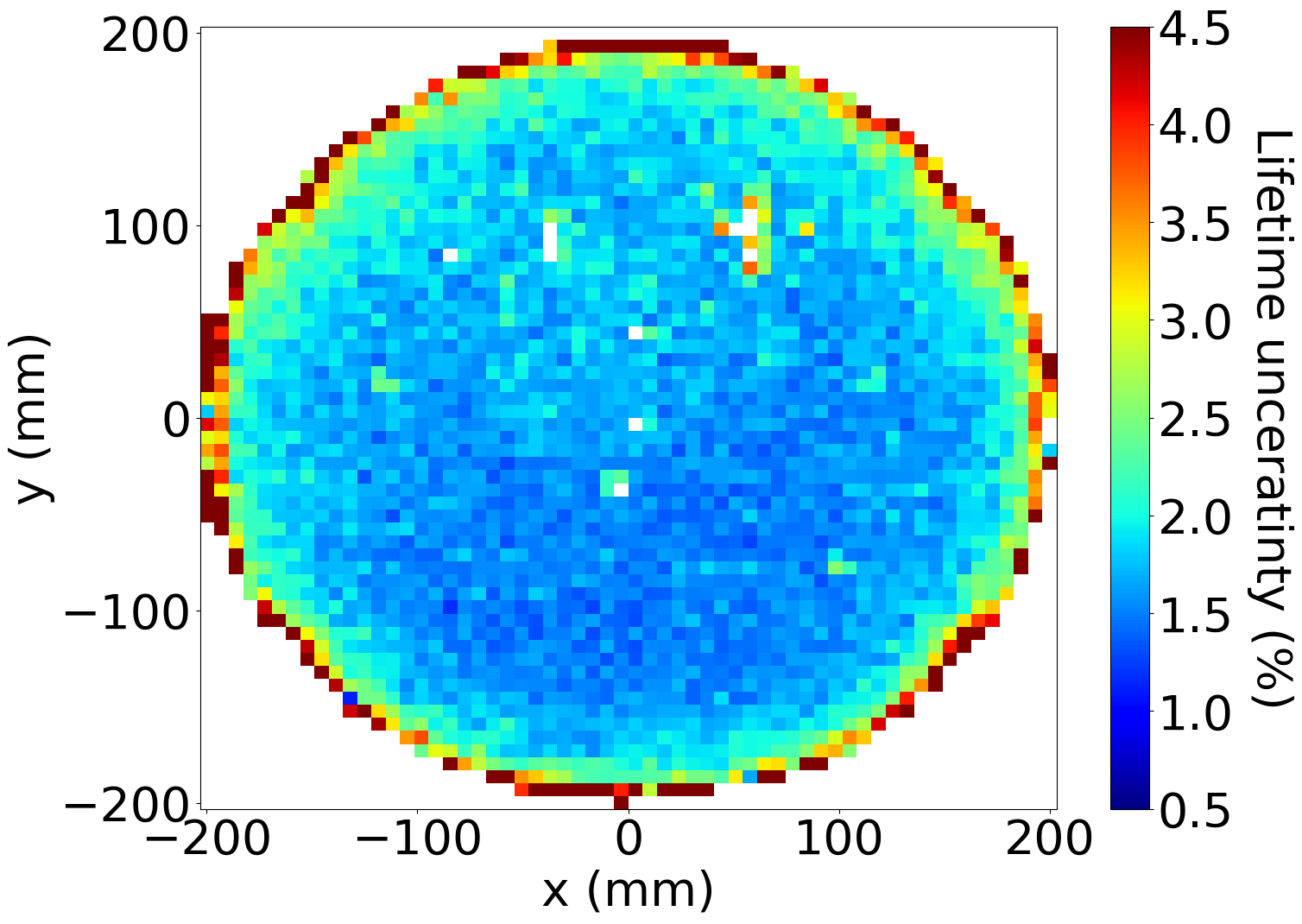}
    \caption{\label{fig:r4734_lt_xymap}  Maps obtained by fitting the lifetime as a function of \XY\ for run 4734. The predicted $e_0$ (left panel) and their uncertainties (right panel) are displayed in the top row, while the lifetimes (left panel) and their uncertainties (right panel) are shown in the bottom row. A clear dependence of the lifetime on \XY\ is observed. The uncertainty in the energy scale is of the order of 0.2 \%, making a sub-dominant contribution to the energy resolution at \KrEnergy. On the other hand, the uncertainty in the lifetime value is of the order of the 1\% and cannot be neglected for the interpretation of the final value of the energy resolution.}
  \end{center}
\end{figure}

\begin{figure}[tbh]
  \begin{center}
    \includegraphics[width=0.4 \textwidth]{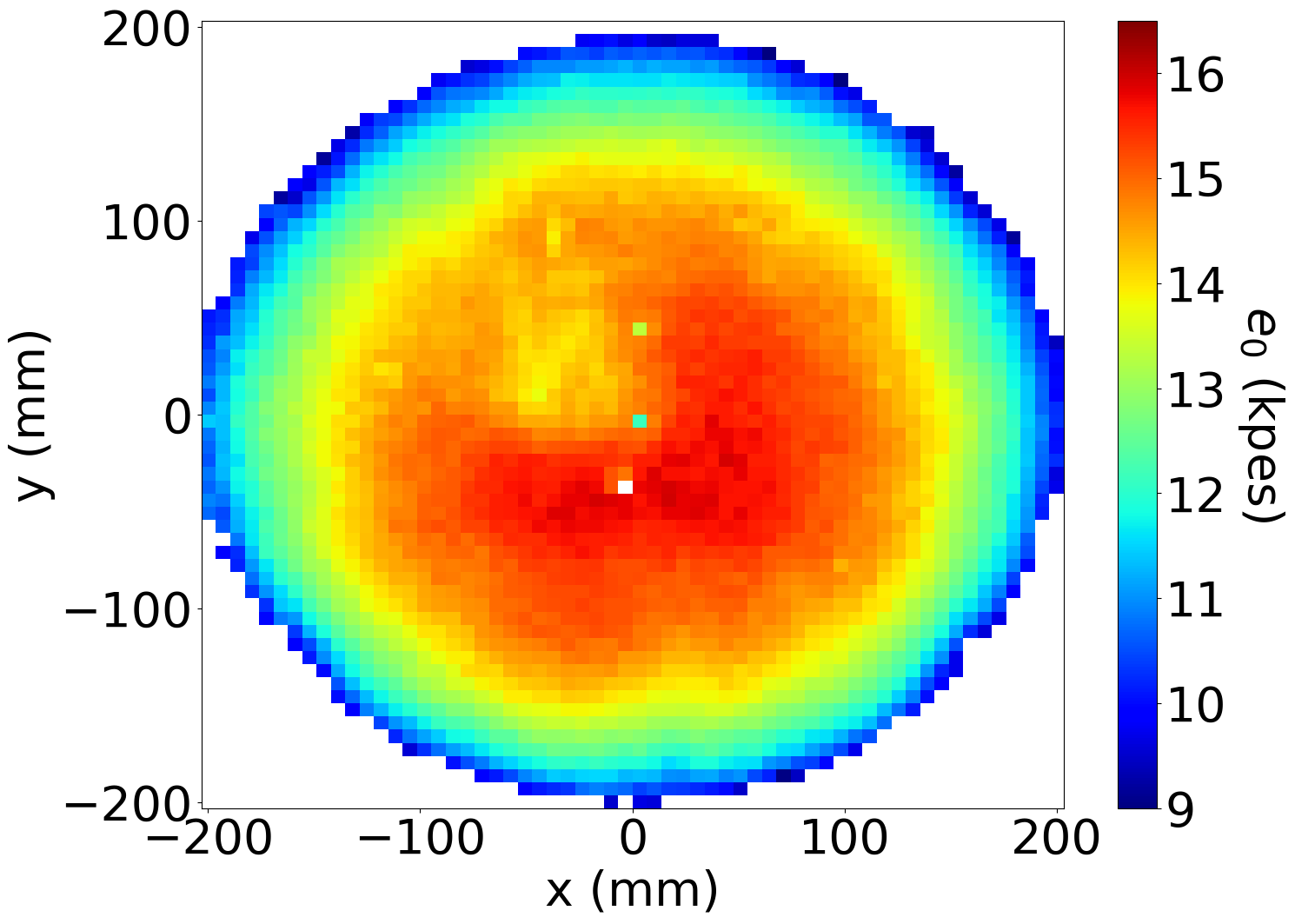}
    \hspace{5mm}
    \includegraphics[width=0.4 \textwidth]{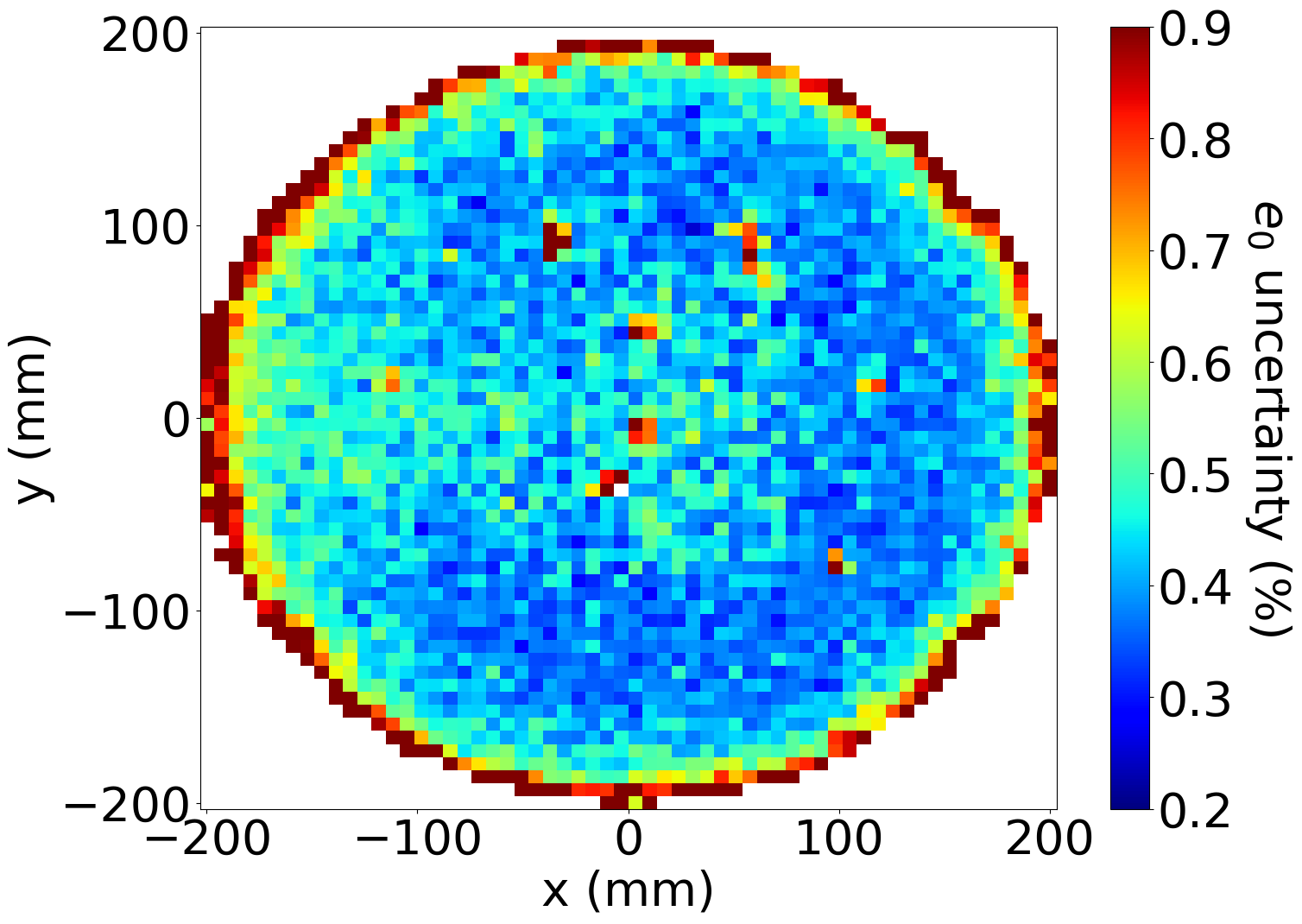}
    \vspace{5mm}
    \includegraphics[width=0.4 \textwidth]{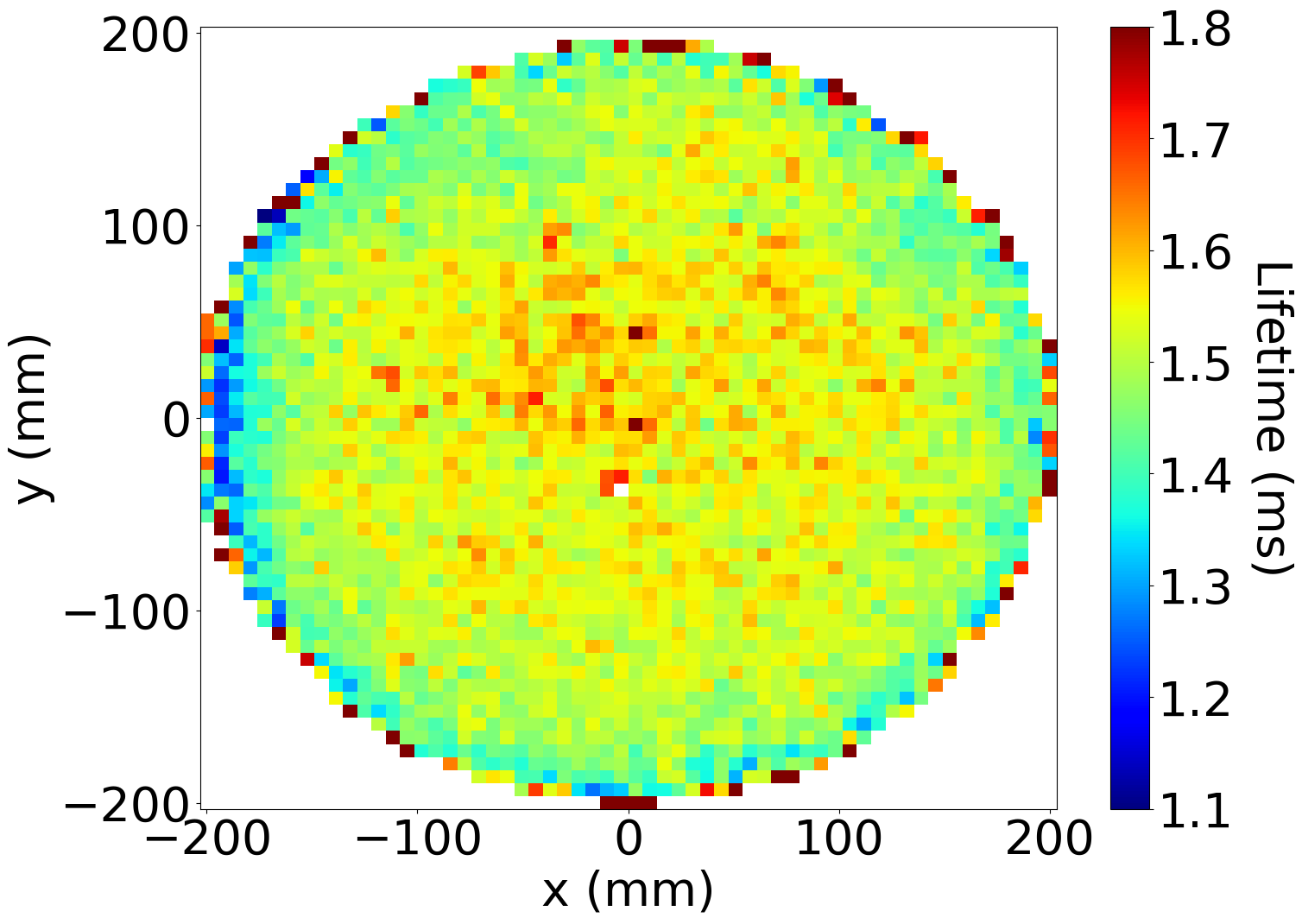}
    \hspace{5mm}
    \includegraphics[width=0.4 \textwidth]{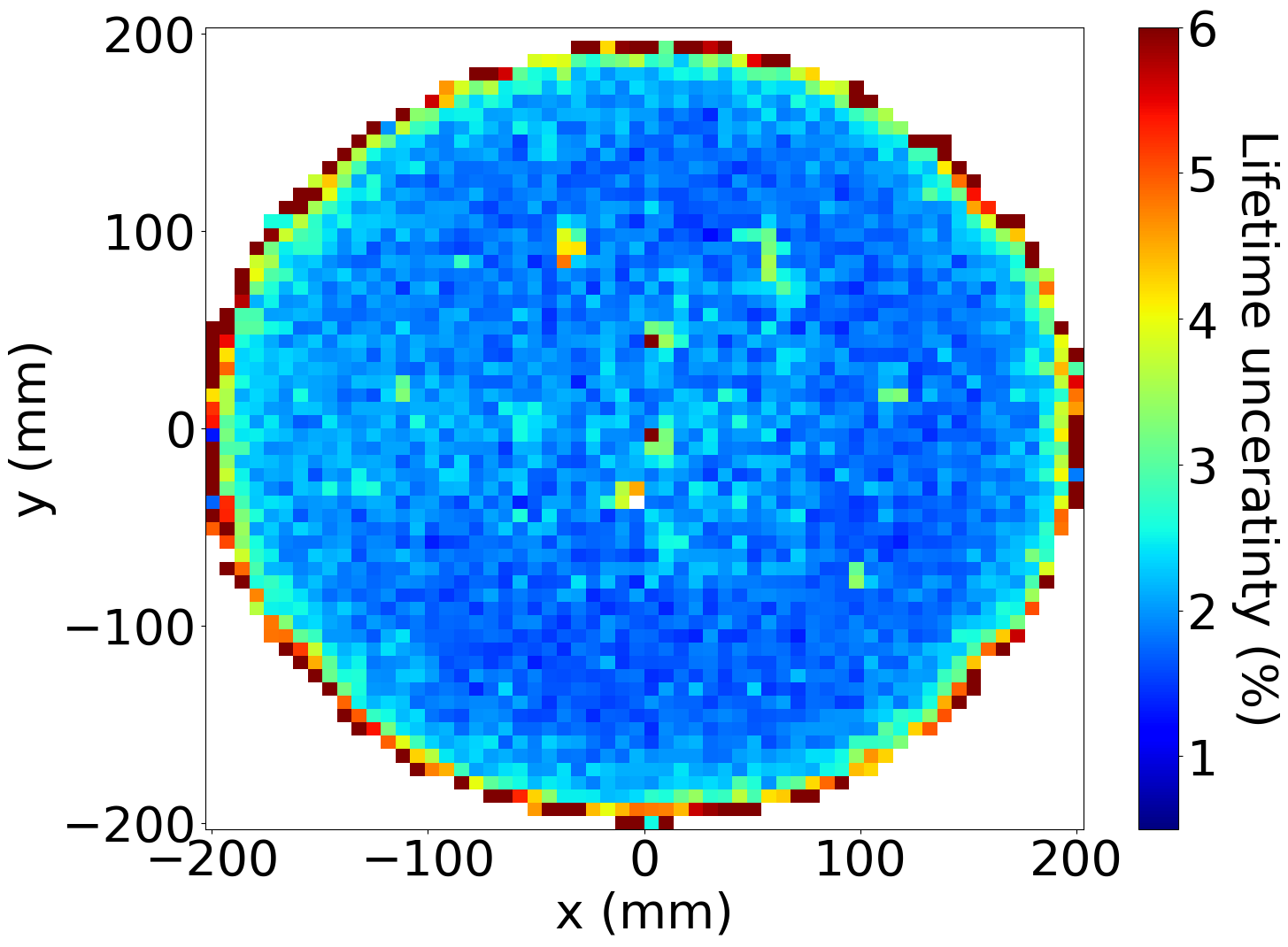}
    \caption{\label{fig:r4841_lt_xymap}  Maps obtained by fitting the lifetime as a function of \XY\ for run 4841. The signal map is statistically compatible with the one obtained with run 4734, while the lifetime map has become homogeneous.}
  \end{center}
\end{figure}

This dependence can be taken into account and corrected for using large-statistics krypton runs to produce a {\em lifetime map}.
The map is built by dividing the chamber in \KryptonLifetimeMapBinsRunII\ \XY\ bins, each of edge \KryptonLifetimeMapBinSizeRunII, and fitting for the lifetime in each bin.
The number of bins is chosen to maximize granularity while still keeping enough data in each bin so that the statistical uncertainties of the fits are small. 

The resulting maps are shown in \fig\ \ref{fig:r4734_lt_xymap} for run 4734.
The fits result in a lifetime parameter and a prediction of $e_0$ for each bin and the corresponding uncertainties.
Thus, the map displayed in the top-left panel is essentially a signal map, where the effect of the lifetime has been factored out, showing the dependence of the event energy on \XY.
The map is rather uniform in the central region, with the exception of a ``crater'' centered around \KryptonEnergyMapPositionCraterRunII\, whose origin we attribute to a few SiPM boards with degraded reflectance, and fall abruptly at large radius, as the solid angle covered by the PMTs falls to zero.
The lifetime map is shown in the bottom-left panel.
A region of longer lifetime (close to \SI{2}{\milli\second}) appears at large positive y, near the top of the chamber
(the average lifetime in the center of the chamber is around \SI{300}{\micro\second} smaller). While this feature is unexpected, study has shown that the flow of the gas through the detector active volume can be turbulent causing regions to have different concentrations of impurities. As can be seen in \Fig\ \ref{fig:r4841_lt_xymap}, which shows the same maps for run 4841, taken at \NewNineBarPressureRunII\ the signal map still shows the crater in the same position, but the lifetime map is uniform, indicating that, as the gas is recirculated and cleaned, these inhomogeneities are gradually removed.%the virtual leaks from ou present during run 4734 have vanished, the gas circulation is more homogeneous, or both. 

The quality of the fits used to extract the lifetime maps as well as the success of their application (see \Fig\ \ref{fig.gausbins} and the clear Gaussian form of the corrected spectra) lead us to believe that the effect is physical and the correction method valid. The crucial point is that, while it is difficult to understand the complex physics that may lead to variable lifetime maps, krypton calibrations permit a correction for those effects.

\section{\label{sec:geometrical_corrections} Refining the energy map}

\begin{figure}[htb]
  \begin{center}
    \includegraphics[width=0.48 \textwidth]{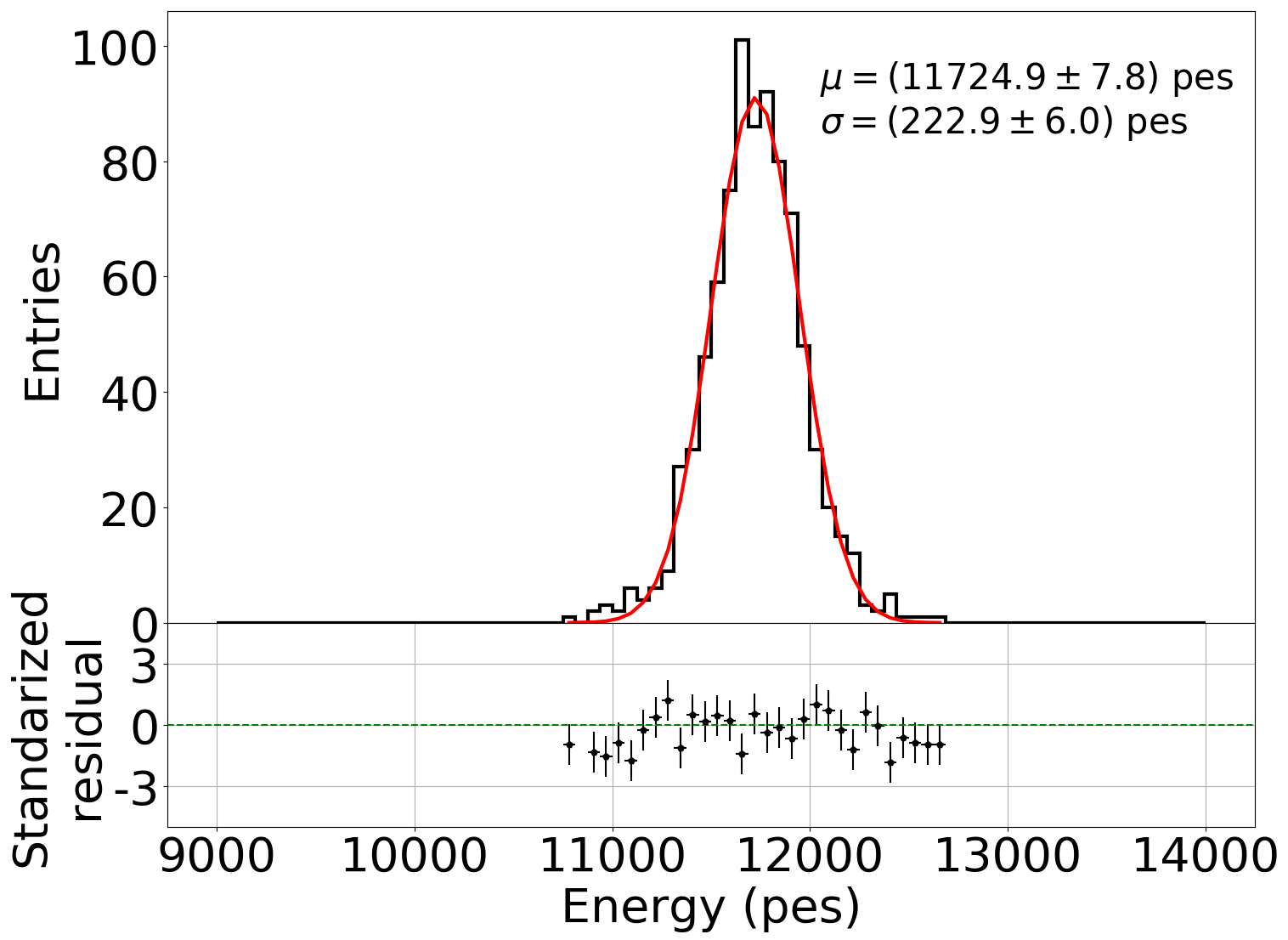}
    \hspace{3mm}
    \includegraphics[width=0.48 \textwidth]{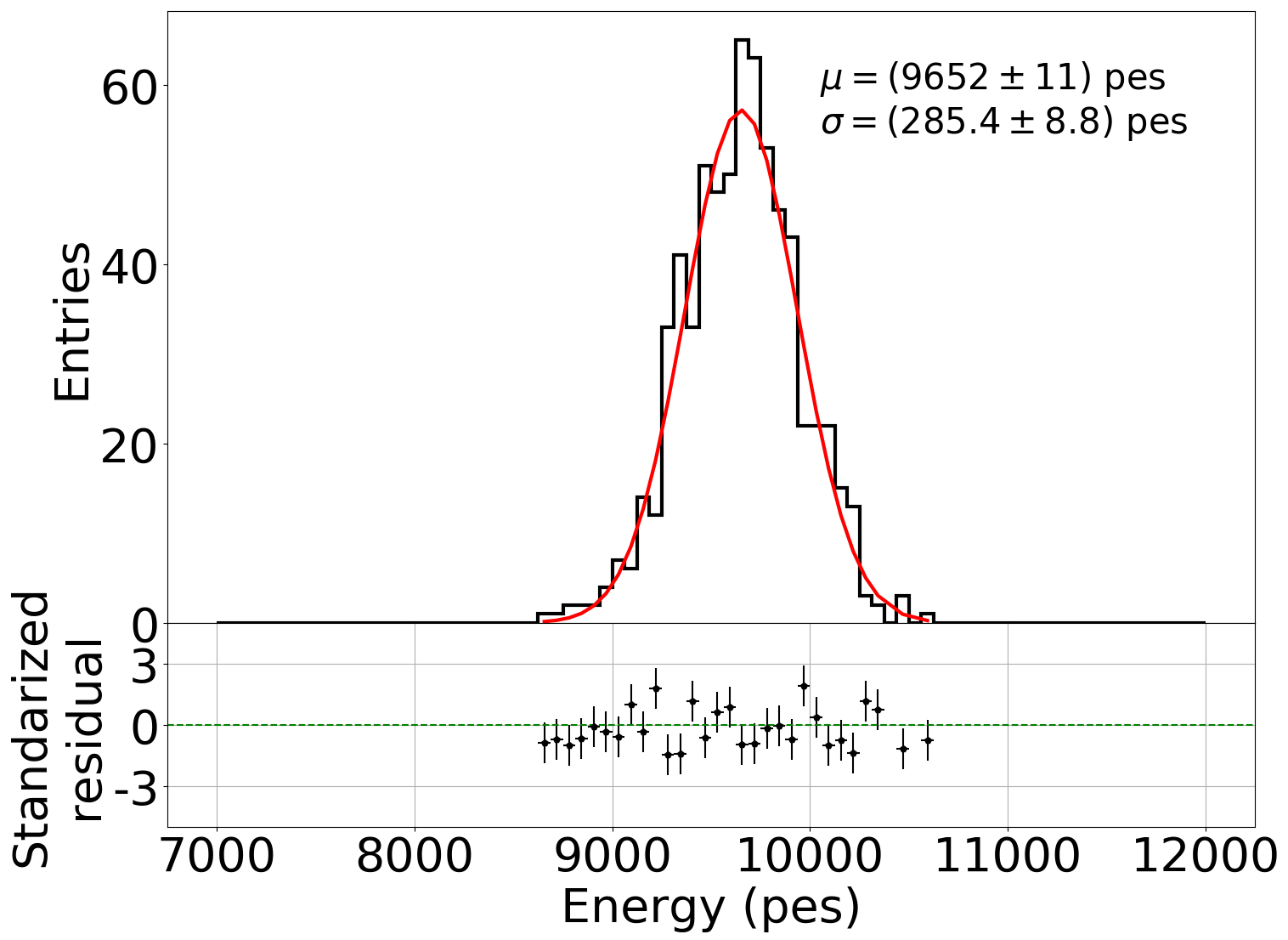}
    \caption{\label{fig.gausbins} Fits to the lifetime-corrected energy for run 4734 in two different regions of the chamber.
             In the region defined by $\rm{x = [0, 10]~mm}   $, $\rm{y = [0, 10]~mm}$ (left panel), the fit yields a mean value for the energy of \SI{11724 +- 8}{\pes}, with $\chitwo$/dof = 0.94, whilst
             in the region defined by $\rm{x = [120, 130]~mm}$, $\rm{y = [120, 130]~mm}$ (right panel), the fit yields a mean value for the energy of \SI{9652 +- 11}{\pes}, with $\chitwo$/dof = 1.03.}
  \end{center}
\end{figure}

Correcting event by event for the fitted lifetimes extracted using the method described in section \ref{sec:lifetime} yields a signal map with the residual variations not related to attachment. The map can be further refined by dividing the \XY\ plane into smaller bins and computing for each bin the sum of the PMT energies corrected by lifetime. The energy in each bin can then be fitted to a Gaussian distribution. An example is shown in \fig\ \ref{fig.gausbins}.  The signal correction factor \FXY\ is simply the inverse of the mean of the gaussian distribution in each \XY\ bin, normalized to a constant factor which can be chosen as the maximum energy bin. 
\begin{figure}[htb]
  \begin{center}
    \subfloat[Data\label{fig:geo_xymap_data}     ]{\includegraphics[width=0.48 \textwidth]{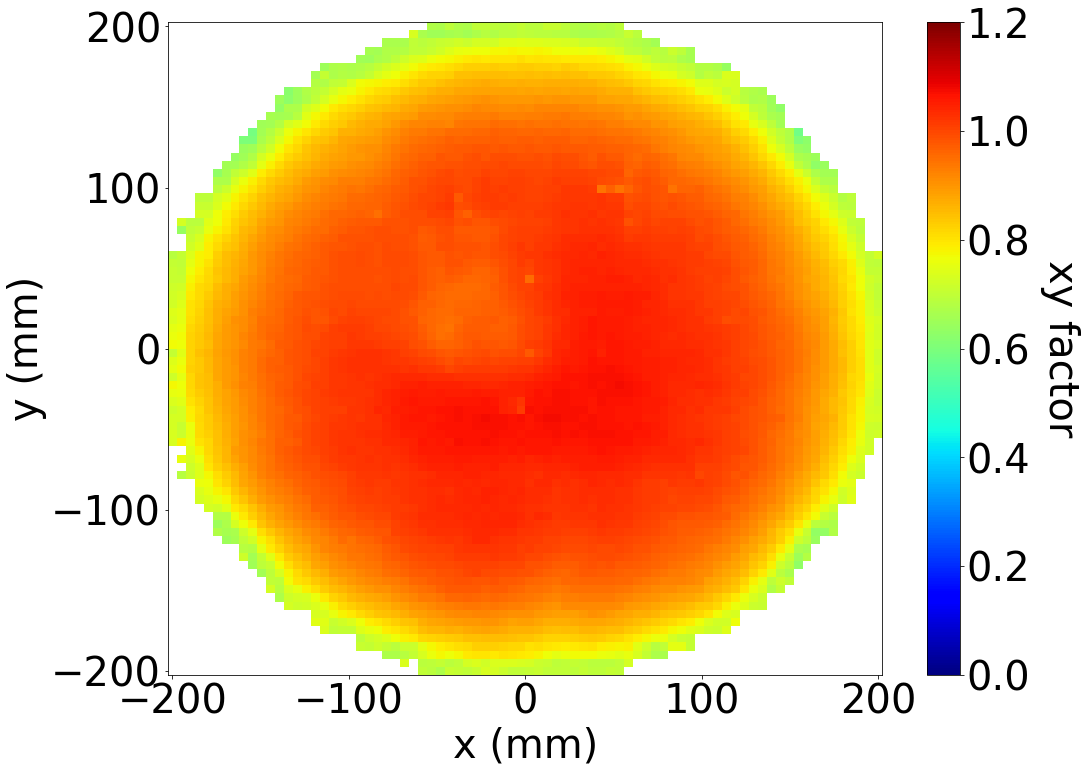}}
    \hspace{3mm}
    \subfloat[Monte Carlo\label{fig:geo_xymap_mc}]{\includegraphics[width=0.48 \textwidth]{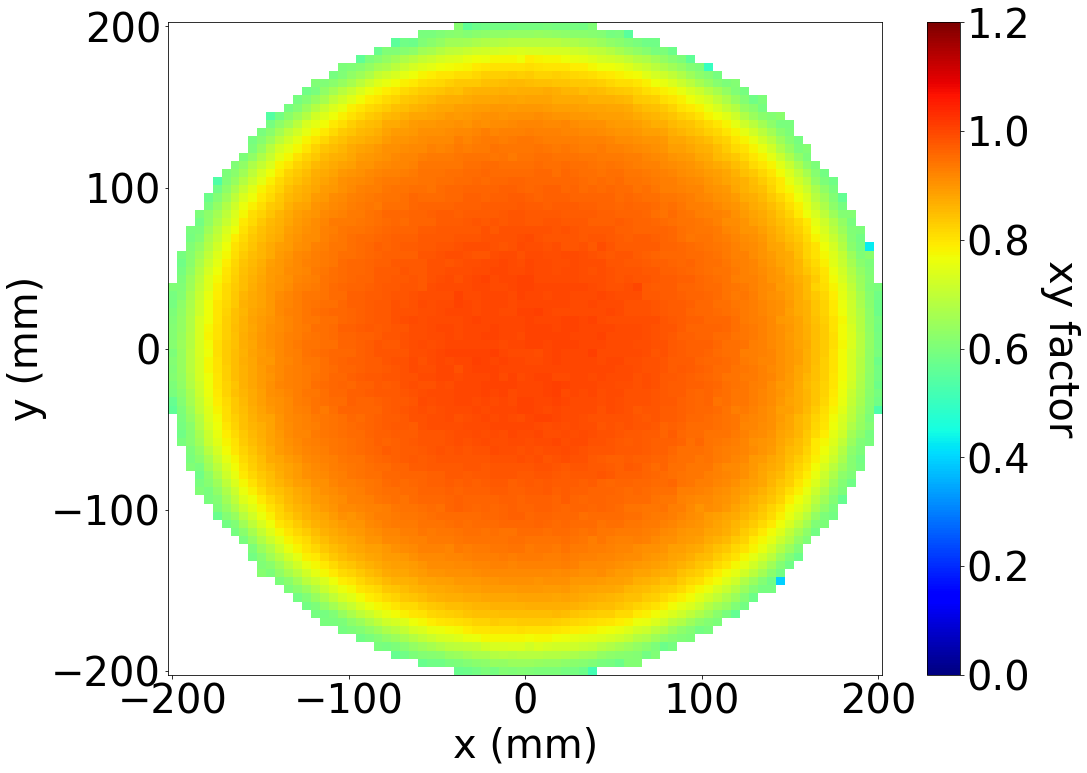}}
    \caption{\label{fig:geo_xymap} Normalized signal map for run 4734 (left panel) and for Monte Carlo data (right panel).}
  \end{center}
\end{figure}

\Fig\ \ref{fig:geo_xymap_data} shows the signal map for run 4734 (the map for 4841 is essentially identical) compared with the signal map computed using Monte Carlo data in \Fig\ \ref{fig:geo_xymap_mc}. Notice that the behavior of the map at large radius, largely due to solid angle effects and edge effects, is well predicted by the Monte Carlo, but not the presence of the crater, which can only be corrected using calibration data. The uncertainties are very small (of the order of \SI{0.3}{\percent}) introducing a small residual error in the energy correction, which for \RII\ is negligible compared with the residual error introduced by the lifetime correction, except at large radius where the angular coverage of the PMTs falls steeply.

\section{\label{sec:energy_resolution} Energy Resolution}

\begin{figure}[tbh]
  \begin{center}
    \includegraphics[width=0.48\textwidth]{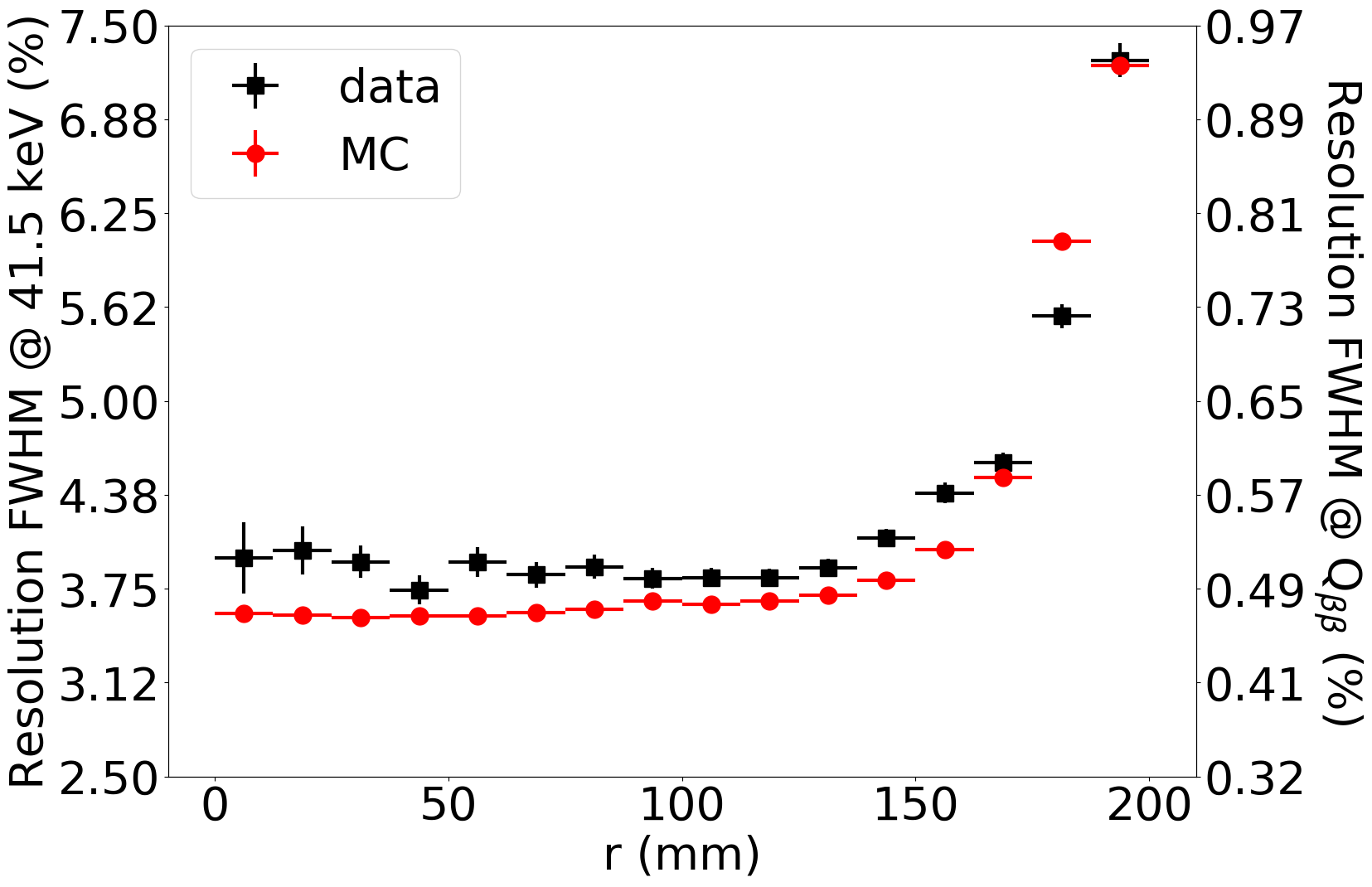}
    \hspace{3mm}
    \includegraphics[width=0.48\textwidth]{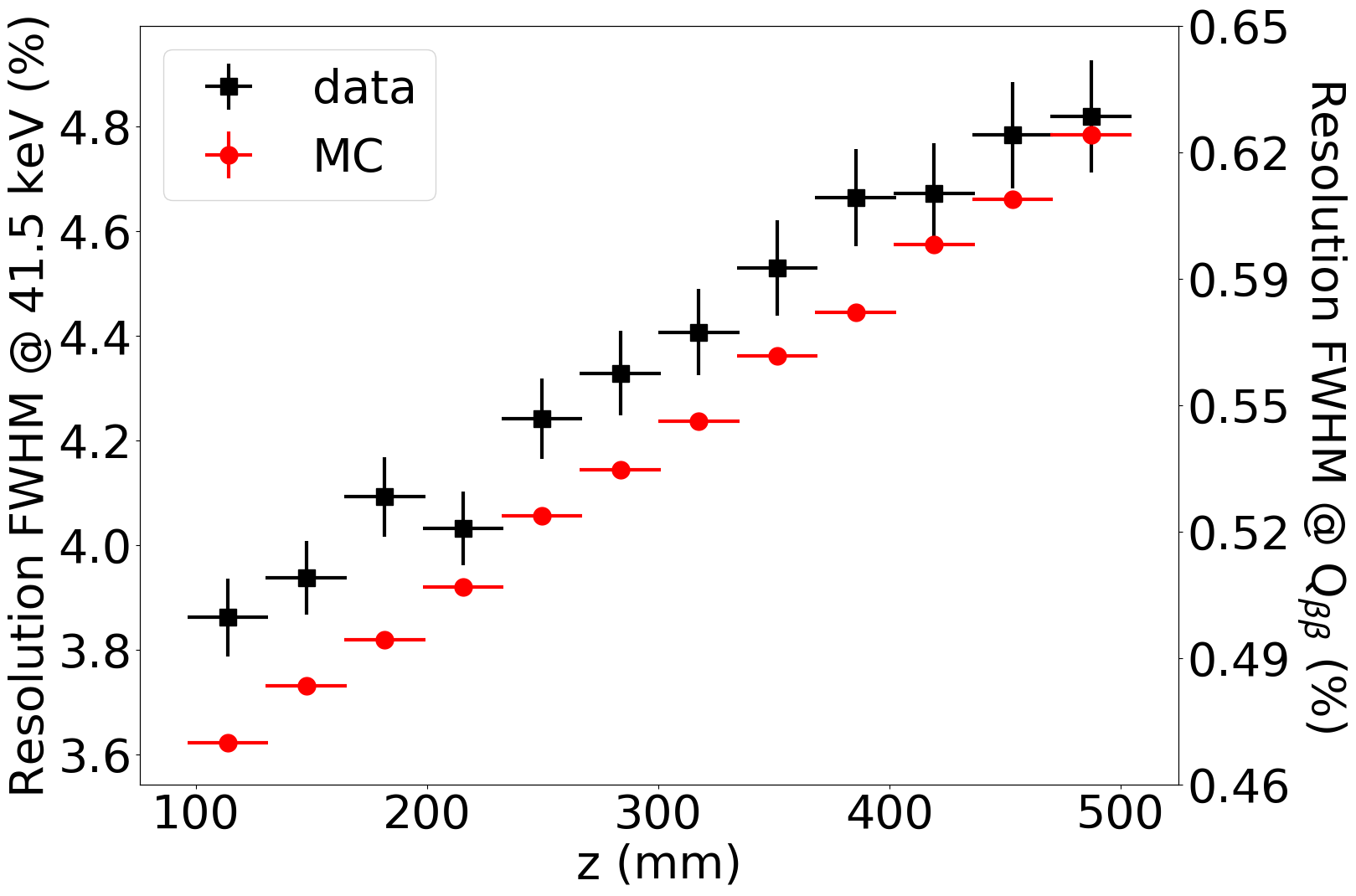}
    \caption{\label{fig:Eres_RZ} Left panel: dependence of the resolution with \R\ for events at short drift time ($\Z < 200$~mm). Right panel: dependence of the resolution with \Z\ for events near the center ($\R < 100$~mm).}
  \end{center}
\end{figure}

To estimate the energy resolution for point-like energy deposits in \NEW, the krypton data are divided in two samples. The {\em correction} sample is used to compute the lifetime and geometry correction maps, which are then applied to the data in the {\em measurement} sample. The corrected signal of the PMT sum is then fitted to estimate the energy resolution. 

Even after corrections, the energy resolution is expected to depend on both the radial and the longitudinal coordinates. The dependence with the radius is related with the decreasing solid angle coverage, which means that PMTs record less light (thus larger fluctuations) for events happening at larger \R. The dependence with \Z\ is related with the loss of secondary electrons caused by attachment. A smaller number of electrons is associated with larger fluctuations and correction factors, which worsens the energy resolution as described in section \ref{sec.hpxe}. 

The left panel of \fig\ \ref{fig:Eres_RZ} shows the dependence of the energy resolution as a function of \R\ for events with $\Z < 200$ mm (black squares data, red circles MC), where it is possible to define 3 regions. A fiducial region up to \R\ $<$ \KrFidVolumeRRunII, where the resolution is roughly flat, at around 4\% FWHM. The resolution stays below 4.5 \%, 
for \R\ $<$\KrExtFidVolumeRRunII, and degrades rapidly for larger radial values. 
Since the total radial coverage of the chamber is \KrTotVolumeRRunII, this implies that an extended fiducial region of acceptable resolution extending up to \KrExtFidVolumeRRunII\ can be defined for physics analysis. The PMT coverage improves as detector radial dimensions increases, since the region of low solid angle coverage corresponds essentially to the last PMT ring.
Taking this into consideration, a considerably smaller reduction in fiducial volume is expected for NEXT-100 since only the last \SIrange{10}{15}{mm} of the total radius need to be removed.

The right panel of \Fig\ \ref{fig:Eres_RZ} shows the dependence of the energy resolution as a function of \Z\ for events with $\R < 100$ mm (black squares data, red circles MC), which degrades with increased drift, although it stays always below 5\% FWHM. The obvious implication is that long lifetimes are a must for TPC detectors striving to achieve excellent energy resolution.  

\Fig\ \ref{fig:Eres_RZ} also shows the energy resolution for MC events (red circles). The effect of the lifetime dependence with \XY, as measured in data, has been also simulated. The agreement between data and MC results is good, indicating that the main dependencies of the resolution are the geometrical effects (lower coverage at a larger radius) and the finite electron lifetime (worse resolution at longer drift times). 

\begin{figure}[tbh]
  \begin{center}
     \includegraphics[width=0.48\textwidth]{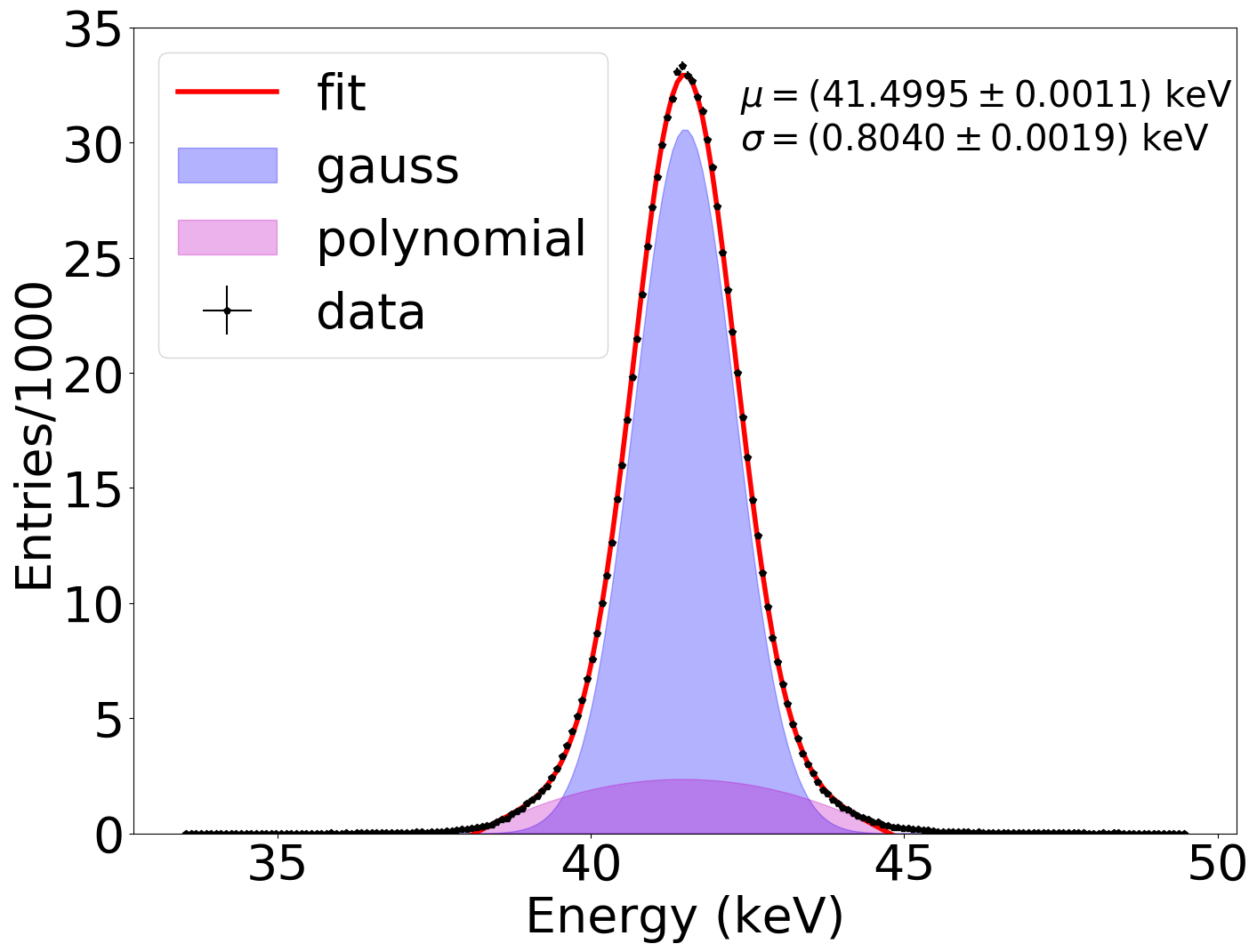}
    \includegraphics[width=0.48\textwidth]{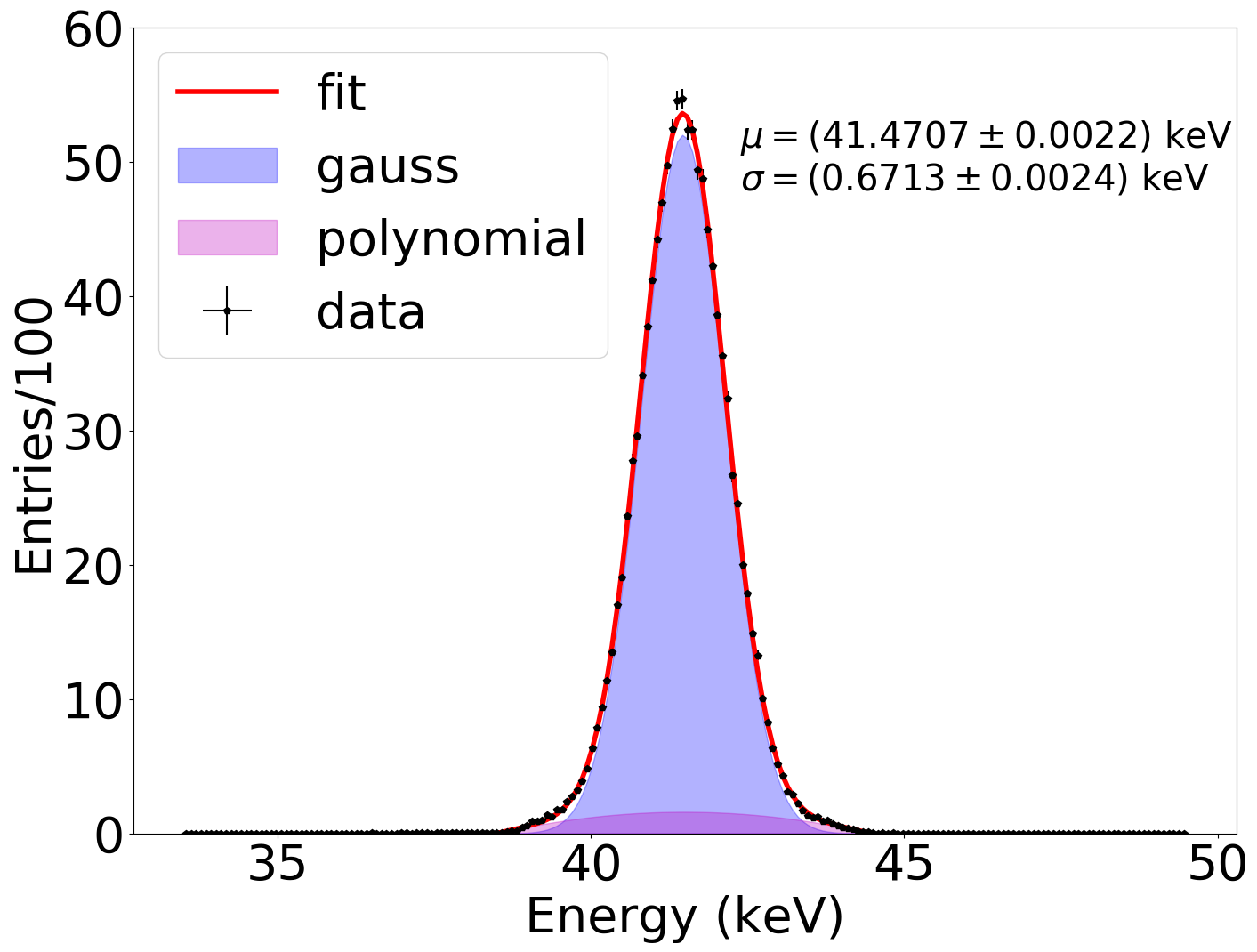}
    \caption{\label{fig:kr_eres_res_4734} Corrected energy distribution for krypton events (left) in the full volume of the \NEW\ TPC, and in a restricted fiducial volume (right), for run 4734.  See text for details.}
  \end{center}
\end{figure}

\begin{figure}[tbh]
  \begin{center}
       \includegraphics[width=0.48\textwidth]{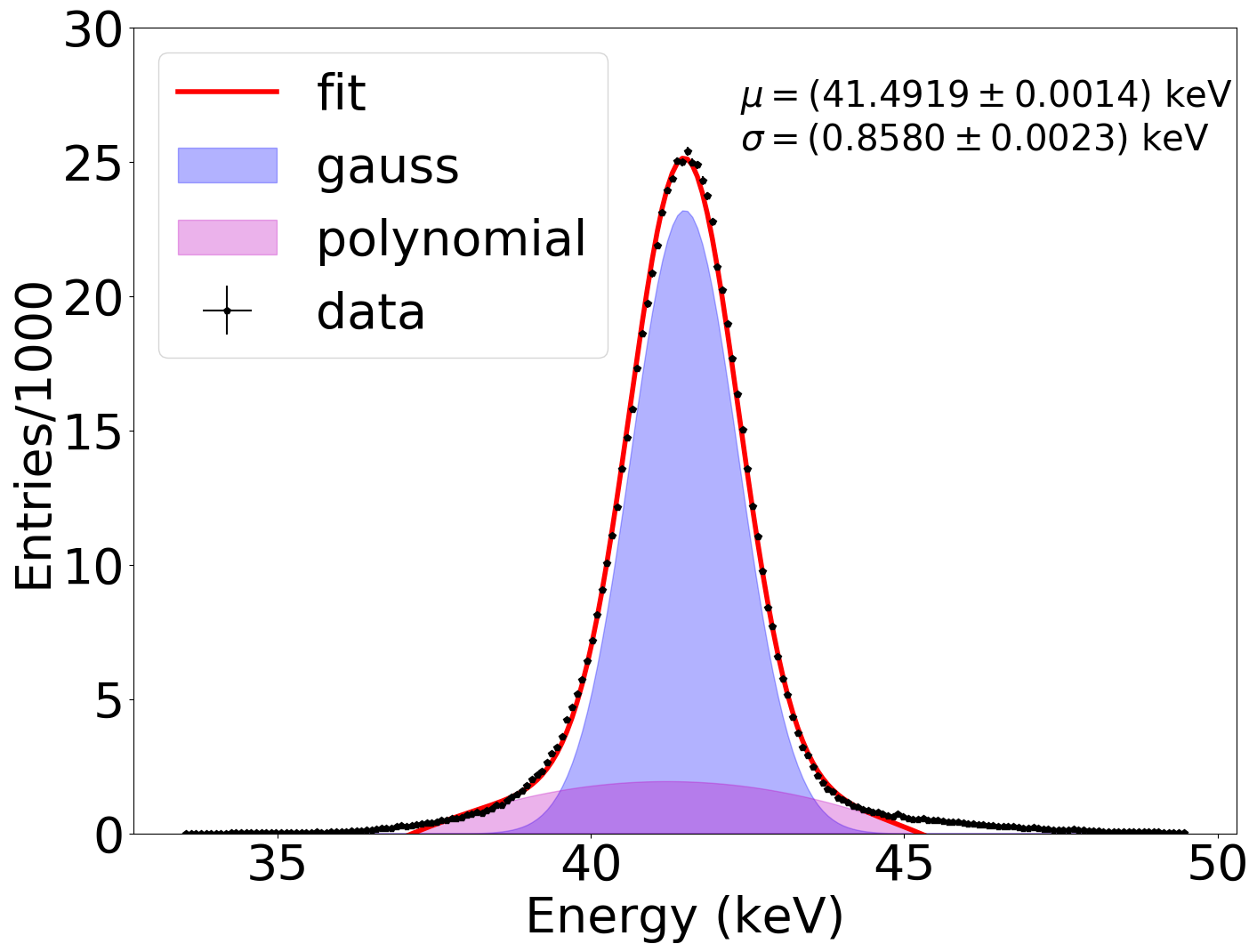}
    \includegraphics[width=0.48\textwidth]{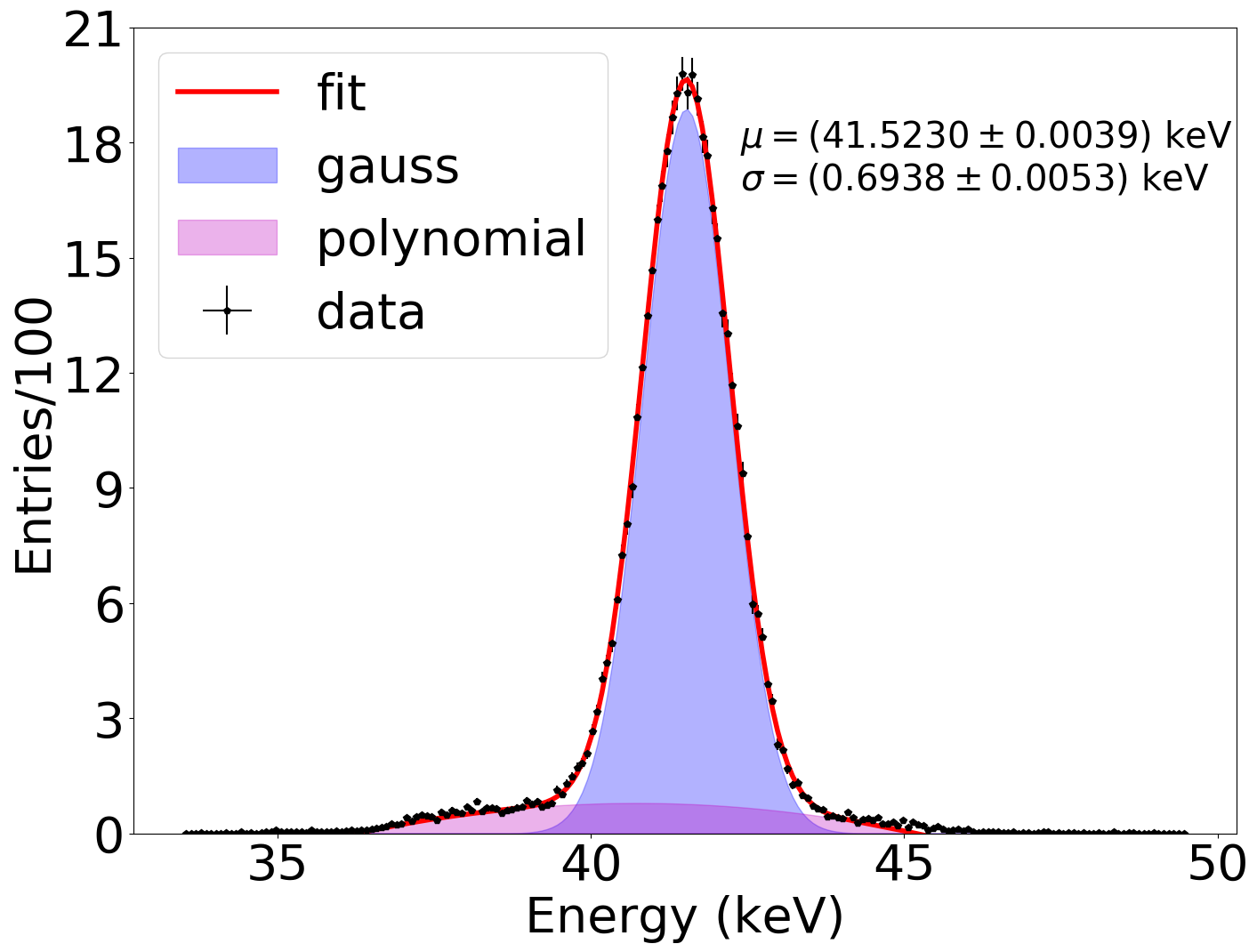}
    \caption{\label{fig:kr_eres_res_4841} Corrected energy distribution for krypton events (left) in the full volume of the \NEW\ TPC, and in a restricted fiducial volume (right), for run 4841.  See text for details.}
  \end{center}
\end{figure}

\Fig\ \ref{fig:kr_eres_res_4734} illustrates the energy resolution measured with run 4734 (at a pressure of \NewSevenBarPressureRunII). The data are fitted to a gaussian plus a 2nd-degree polynomial to take into account tails due to residual background events (small energy deposits or \Kr{83m} decays with wrong \so\ identification). The fit yields an energy resolution of \ResolutionKrFullFourSevenThreeFour\ FWHM in the full \NEW\ volume (left panel). A naive
\SQRE\ extrapolation to \Qbb\ yields \ResolutionKrFullFourSevenThreeFourQbb. The fit in the right panel corresponds to the data contained in a fiducial region defined by a radius smaller than 
\KrFidVolumeRRunII\ and \Z\ smaller than \KrFidVolumeZRunII. The radial cut ensures optimal geometrical coverage and the \Z\ cut minimizes the residual errors due to lifetime fluctuations, which increase with \Z. The fit yields \ResolutionKrFidFourSevenThreeFour, extrapolating to \ResolutionKrFidFourSevenThreeFourQbb\ at \Qbb. This value is reasonably close to the best resolution expected in \NEW\ (figure \ref{fig.yield}), confirming the excellent capabilities of the technology and the good working conditions of the chamber.

The same procedure is applied to run 4841 (at a pressure of \NewNineBarPressureRunII) in \Fig\ \ref{fig:kr_eres_res_4841}.
The fit yields an energy resolution of \ResolutionKrFullFourEightFourOne\ FWHM in the full \NEW\ volume (left panel).
A naive \SQRE\ extrapolation to \Qbb\ yields \ResolutionKrFullFourEightFourOneQbb.
The fit in the right panel corresponds to the data contained in the fiducial region defined above.
The fit yields \ResolutionKrFidFourEightFourOne, extrapolating to \ResolutionKrFidFourEightFourOneQbb\ at \Qbb, similar to the values obtained for run 4734, and confirming that resolution for point-like energy deposits scales well with pressure. Higher energy depositions are studied and a more complete study of the extrapolation is made in another analysis \cite{Renner:2018ttw}.

\Tab~\ref{tab:KrSystematics} summarizes the contributions to the systematic uncertainty of the energy resolution considering the full and fiducial volumes for both run 4734 and run 4841. The main systematic errors come from the lifetime and geometrical corrections, the fit range and the model used to describe the tails of the energy distribution. The systematic uncertainties of the lifetime and geometrical corrections have been estimated by measuring the variation of the energy when those factors are shifted by $\pm 1 \sigma$ around their optimal value. The systematic uncertainty associated to the bin size has been estimated as the maximum difference of the resolution when varying the bin size, within a sensible range, of the energy spectrum histogram. In order to estimate the uncertainties related to the fit model, we have considered different functions to describe the tails of the energy distribution and used the maximum difference of the resolution among those models and fits that resulted in an acceptable goodness of the fit. The total systematic uncertainty, adding the different contributions in quadrature, is 0.324 \% (0.112 \%) for run 4734 and 0.246 \% (0.148 \%) for run 4841 for the full (fiducial) volume. 

We get a final estimate of the energy resolution in the full volume of \linebreak \ResolutionKrFullFourSevenThreeFourWithSystematics\ FWHM (\ResolutionKrFullFourSevenThreeFourQbbWithSystematics \linebreak FWHM at \Qbb) for the \NewSevenBarPressureRunII\ run and \ResolutionKrFullFourEightFourOneWithSystematics\ FWHM \linebreak (\ResolutionKrFullFourEightFourOneQbbWithSystematics\ FWHM at \Qbb) for the \NewNineBarPressureRunII\ run.

\begin{table}
    \centering
    \begin{tabular}{|c|cc|cc|}
        \cline{2-5}
        \multicolumn{1}{c|}{} & \multicolumn{2}{c|}{Run 4734} & \multicolumn{2}{c|}{Run 4841} \\ \cline{2-5}
        \multicolumn{1}{c|}{} &  Full  & Fiducial &  Full & Fiducial \\ \hline
        Lifetime map          &  0.044 \% &   0.008 \% & 0.100 \% & 0.016 \% \\ \hline
        Geometry map          &  0.015 \% &   0.018 \% & 0.030 \% & 0.031 \% \\ \hline
        Fit range             &  0.265 \% &   0.064 \% & 0.143 \% & 0.011 \% \\ \hline
        Histogram binning     &  0.002 \% &   0.003 \% & 0.002 \% & 0.011 \% \\ \hline
        Background model      &  0.182 \% &   0.090 \% & 0.172 \% & 0.143 \% \\ \hline \hline
        Total                 &  0.324 \% &   0.112 \% & 0.246 \% & 0.148 \% \\ \hline
    \end{tabular}
    \caption{Main contributions to the systematic uncertainty in the measurement of the energy resolution at 41.5 keV.}
    \label{tab:KrSystematics}
\end{table}

\section{\label{sec:summary} Summary}

The  \NEW\ detector has been calibrated using large samples of \Kr{83m} decays taken near the end of the long calibration run (\RII) acquired in 2017. Two large data samples have been analyzed for this paper, run 4734 taken at a pressure of \NewSevenBarPressureRunII, and run 4841 taken at \NewNineBarPressureRunII. The average lifetimes of the chamber were around \SI{1.8}{ms} and \SI{1.4}{ms}, respectively. In run 4734 a clear dependence of the lifetime with the transverse position \XY\ is observed, while run 4841 shows a constant lifetime in all the chamber. We observe that the energy map is in very good agreement with the Monte Carlo prediction, although it shows an unexpected small region of lower response near the center, on the top left side of the detector, presumably due to degraded reflectance in some SiPM boards. The effect of lifetime and solid angle is taken into account by correcting the data with lifetime and energy maps. 
 
%The measured energy resolution for point-like energy deposits in \NEW\ at \NewSevenBarPressureRunII\ (run 4734) is \ResolutionKrFullFourSevenThreeFourWithSystematics\ (which extrapolates \SQRE\ to \linebreak \ResolutionKrFullFourSevenThreeFourQbbWithSystematics) in the full chamber and
%\ResolutionKrFidFourSevenThreeFourWithSystematics\ (\ResolutionKrFidFourSevenThreeFourQbbWithSystematics) in a fiducial region
%($\R < \KrFidVolumeRRunII, \Z < \KrFidVolumeZRunII$) chosen to minimize the effect of lower solid angle coverage and large lifetime corrections. The energy resolution we obtain is remarkably close to the limit value for these conditions: \NewIntrinsicEnergyResolutionRunII. The energy resolution is essentially the same at \NewNineBarPressureRunII\ (run 4841). The results show the robustness of the technique to calibrate the \NEW\ detector as well as the excellent energy resolution characteristic of high pressure xenon chambers.

The measured energy resolution FWHM for point-like energy deposits in \NEW\ at \NewSevenBarPressureRunII\ (run 4734) is \ResolutionKrFullFourSevenThreeFourWithSystematics\ in the full volume and \linebreak \ResolutionKrFidFourSevenThreeFourWithSystematics\ in a fiducial region defined by $\R < \KrFidVolumeRRunII$ and $\Z < \KrFidVolumeZRunII$, chosen to minimize the effect of lower solid angle coverage and large lifetime corrections.
Assuming a \SQRE\ extrapolation we find, respectively, \ResolutionKrFullFourSevenThreeFourQbbWithSystematics\ and \ResolutionKrFidFourSevenThreeFourQbbWithSystematics.
\ (\ResolutionKrFidFourSevenThreeFourQbbWithSystematics).
The energy resolution we obtain is remarkably close to the limit value for these conditions: \NewIntrinsicEnergyResolutionRunII.
The energy resolution is essentially the same at \NewNineBarPressureRunII\ (run 4841).
The results show the robustness of the technique to calibrate the \NEW\ detector as well as the excellent energy resolution characteristic of high pressure xenon chambers.

\acknowledgments
The NEXT Collaboration acknowledges support from the following agencies and institutions: the European Research Council (ERC) under the Advanced Grant 339787-NEXT; the European Union's Framework Programme for Research and Innovation Horizon 2020 (2014-2020) under the Marie Sk\l{}odowska-Curie Grant Agreements No. 674896, 690575 and 740055; the Ministerio de Econom\'ia y Competitividad of Spain under grants FIS2014-53371-C04, the Severo Ochoa Program SEV-2014-0398 and the Mar\'ia de Maetzu Program MDM-2016-0692; the GVA of Spain under grants PROMETEO/2016/120 and SEJI/2017/011; the Portuguese FCT and FEDER through the program COMPETE, projects PTDC/FIS-NUC/2525/2014 and UID/FIS/04559/2013; the U.S.\ Department of Energy under contract numbers DE-AC02-07CH11359 (Fermi National Accelerator Laboratory), DE-FG02-13ER42020 (Texas A\&M), DE-SC0017721 (University of Texas at Arlington), and DE-AC02-06CH11357 (Argonne National Laboratory); and the University of Texas at Arlington. We also warmly acknowledge the Laboratorio Nazionale di Gran Sasso (LNGS) and the Dark Side collaboration for their help with TPB coating of various parts of the NEXT-White TPC. Finally, we are grateful to the Laboratorio Subterr\'aneo de Canfranc for hosting and supporting the NEXT experiment.

\bibliographystyle{pool/JHEP}
\bibliography{pool/NextRefs}

\end{document}